\documentclass[a4paper,11pt]{extarticle}

	\usepackage{enumerate,hyperref,color,siunitx}
	\usepackage[symbol]{footmisc}
	\usepackage{enumitem,layouts,calc}
	\usepackage{siunitx}
	\usepackage{mathtools}
	\usepackage{blkarray}
	\usepackage{placeins}
	
	\usepackage[]{cite}
	\newcommand{\killpunct}[1] 
	
	\usepackage{bm,amsmath,amssymb,physics}
	
	\usepackage[]{graphicx}
	\usepackage[font=small,labelfont=bf,figurename=Figure,%
		labelsep=period]{caption}

	\usepackage{multirow,tabularx}
	\usepackage{algorithm,algpseudocode}
	
	\usepackage[margin=1in]{geometry}
	\usepackage{float,pdflscape,authblk,setspace,lineno}	

	\newenvironment{keywords}
    {\footnotesize \textbf{Keywords:} }
    { \vspace{0.2cm} }

\usepackage[capitalize]{cleveref}
\usepackage{chngcntr}

\usepackage{float}

\begin{document}

	\title{Phenotypic heterogeneity in temporally fluctuating environments}


	\author[1,2*]{Alexander P Browning}
	\author[3]{Sara Hamis}
	
	\affil[1]{School of Mathematics and Statistics, University of Melbourne, Melbourne, Australia}
	\affil[2]{Mathematical Institute, University of Oxford, Oxford, United Kingdom}
	\affil[3]{Department of Information Technology, Uppsala University, Uppsala, Sweden}


		\maketitle

		\vfill
		\begin{abstract}
			\noindent Many biological systems regulate phenotypic heterogeneity as a fitness-maximising strategy in uncertain and dynamic environments. Analysis of such strategies is typically confined both to a discrete set of environmental conditions, and to a discrete (often binary) set of phenotypes specialised to each condition.  In this work, we extend theory on both fronts to encapsulate a potentially continuous spectrum of phenotypes arising in response to environmental fluctuations that drive changes in the phenotype-dependent growth rate. We consider two broad classes of stochastic environment: those that are temporally uncorrelated (modelled by white-noise processes), and those that are correlated (modelled by Poisson and Ornstein-Uhlenbeck processes).  For tractability, we restrict analysis to an exponential growth model, and consider biologically relevant simplifications  that pertain to the timescale of phenotype switching relative to fluctuations in the environment.  These assumptions yield a series of analytical and semi-analytical expressions that reveal environments in which phenotypic heterogeneity is evolutionarily advantageous. 
		\end{abstract}
				\vfill
		\vfill

	\footnotetext[1]{Corresponding author: alex.browning@unimelb.edu.au}


\section{Introduction}

Heterogeneity is ubiquitous to biology \cite{Elsasser.1984uy,Wilkinson:2009,Ackermann.2015,Gough.2016}. Many bacteria strains, for example, employ phenotypic heterogeneity or \textit{persistence} as a bet-hedging strategy against environmental variability \cite{Bigger:1944,Balaban:2004,Jong:2011}. Similar strategies are also employed in cancer, a disease in which persistence and tolerance is now recognised as an important contributor to drug resistance \cite{Turke.2010,Sun.2024}. Such hedging strategies yield not only an fitness advantage to cell populations \cite{Kussell.2005}, but are thought to be essential for long-term survival by safeguarding against extinction \cite{Wolf.2005,Thakur.2023}.  

The degree to which populations bet-hedge is thought to be strongly associated with environment. For instance, so-called high persistence mutants of \textit{Escherichia coli} carry an evolutionary advantage in more frequently stressful environments, but can be outcompeted in less stressful conditions \cite{Kussell.2005,Browning.202115}. Complicating the study of bet-hedging is fundamental uncertainty around the environments under which a given organism evolved. Research efforts focus, therefore, on the emergence of evolutionary stable strategies in response to experimentally (or theoretically) prescribed environments \cite{Beaumont.2009,Bergh.2016}. Most typical are environmental models that comprise periodic switching between a finite number of conditions \cite{Thattai:2004,Kussell.2005,Belete.2015} : the presence or absence of an drug, antibiotic, or other stressor, for instance, or the relative prevalence of various nutrient sources \cite{Lambert.2014}.

Driven in part by advances in experimental capability, recent years have seen the emergence of a wider spectrum of arguably more realistic experimental and theoretical models of environmental variability \cite{Letten.2023}. In addition to stochastic analogues of the aforementioned discrete models of environment (typically driven by inhomogeneous Poisson processes)  \cite{Wolf.2005,Acar.2008,Muller.201303s,Taitelbaum.2020,Hernandez-Navarro.2023} , recent theoretical work enables interrogation of bet-hedging in environments driven by more general stochastic processes \cite{Browning.202115,Hidalgo.2015,Liu.2019,Taitelbaum.2023,Sireci.2023}. Our recent work \cite{Browning.202115}, among others \cite{Hidalgo.2015,Muller.201303s}, draws on the similarities between cellular and financial hedging to demonstrate that phenotypic heterogeneity can be evolutionarily advantageous in environments driven by white-noise processes that yield growth rates with constant mean.  These works also study what we refer to as \textit{continuously fluctuating environments} in which environmental change is continuous, such in environments described by Ornstein-Uhlenbeck processes or generalisations thereof.  There is, however, little general theory available to elucidate bet-hedging in arbitrarily fluctuating environments.

Microorganisms regulate phenotypic variability through gene expression such that otherwise genetically identical individuals exhibit differences in traits \cite{Ackermann.2015}. Bet-hedging bacteria, for example, exhibit a genetically-determined switching strategy that regulates small subpopulations of \textit{persister} phenotypes that protect the population from future environmental stress \cite{Balaban:2004,Kussell.2005}. Such quiescent phenotypes are typically characterised by a metabolism that differs from regular proliferative bacteria: this manifests as both lower proliferation rates (in growing conditions) and lower death rates (in stressful conditions) \cite{Balaban:2004,Kussell.2005v2,Williams.2011}. Persisters can be formed through both active switching (i.e., in response to environmental stressors through stress response hormones including (p)ppGpp \cite{Day.2016,Irving.2021,Groot.2023}) and passive switching (i.e., through environment-independent phenotype switching \cite{Balaban:2004}). Passive switching may be regulated through stochastic gene expression, whereby sufficiently high levels of a persistence gene may correspond to the quiescent state.  This expression-threshold theory underscores the overwhelmingly common view of heterogeneity as a discrete (and often binary) set of otherwise homogeneous subpopulations \cite{Verstraete.2021,Fridman.2014}. Subpopulation level variation is typically not considered in the analysis of bet-hedging strategies outside a small number of largely theoretical works \cite{Svardal.2011,Levien.2020,Mateu.2021,Kerr.2022}.  While models that capture a continuous distribution of phenotypes are widely studied in the mathematical literature (typically in the context of cancer plasticity \cite{Chisholm.2015,Browning.2024dzf}), there are few studies that investigate such \textit{continuous phenotypic heterogeneity} in stochastically fluctuating environments \cite{Bodova.2021}. 

Theoretical models of bet-hedging range in scale from intracellular studies of the pathways responsible for persister regulation \cite{Garcia:2015}, to high-level studies of populations that regulate heterogeneity through phenotype switching \cite{Kussell.2005v2}. At the population-level, works commonly study the evolutionarily stable strategy (ESS), or fitness-maximising strategy, under a prescribed growth model coupled to an environmental process \cite{Tuljapurkar.1990,Metz.2008,Bukkuri.2023s28}. Growth is typically modelled to be exponential, representative of batch- or continuous-culture experiments, where cell concentrations are kept sufficiently low that intercellular competition---which may obfuscate or otherwise alter the evolutionary dynamics---is negligible. Study of the ESS then corresponds to a stochastic optimisation problem in the Malthusian fitness or, in the case that organisms are able to actively react to the environment, a stochastic optimal control problem \cite{Browning.202115,Mononen.2023}. 

	\begin{figure}[!t]
		\centering
		\includegraphics[]{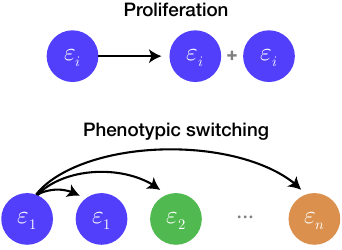}
		\caption[Fig 1]{\textbf{Population dynamics model.}   We assume that cell proliferation is symmetric (top), while phenotypic heterogeneity is regulated through switching where, when a switching event is triggered, each cell is reassigned a phenotype according to a discretely or continuously structured distribution denoted $\hat{p}(\varepsilon)$.}
		\label{fig1}
	\end{figure}

In this work, we develop mathematical theory to study the environmental conditions in which bet-hedging is advantageous. Throughout, we consider a Malthusian-like measure of fitness given by the expected per-capita growth rate \cite{Tuljapurkar.1990,Metz.2008,Williams.2011}
	\begin{equation}\label{payoff}
		\varphi = \mathbb{E}\left(\dfrac{1}{n}\dv{n}{t}\right) = \mathbb{E}\left(\dv{\log n}{t}\right),
	\end{equation}
where $n(t)$ denotes the overall size (or concentration) of a phenotypically structured population. We make the standard assumption that phenotypic traits are heritable through symmetric exponential division, and that phenotypic heterogeneity is regulated through a phenotype switching strategy (\cref{fig1}) \cite{Balaban:2004,Kussell.2005}. Evolutionary stable strategies are, therefore, switching strategies that maximise \cref{payoff}. In \cref{sec_whitenoise}, we consider a fitness maximising phenotype switching strategy in an environment driven by a pair of temporally uncorrelated white-noise processes.  We then, in \cref{sec_continuous}, extend the analysis: first to temporally correlated environments driven by an inhomogeneous Poisson process, and then to a family of environments driven by an Ornstein-Uhlenbeck process such that growth rate changes are continuous.  In all but the simplest of cases, the calculation of population fitness (\cref{payoff}) is  both numerically and analytically intractable. Thus, we make progress by studying a series of biologically relevant limiting cases based around the relative time-scales of environmental fluctuations and phenotypic-switching. This approach allows us to determine sufficient conditions for parameter regimes in which discrete and continuous heterogeneity is advantageous.


\section{Growth in environments driven by temporally uncorrelated white-noise}\label{sec_whitenoise}

	We first consider a cell population subject to environmental fluctuations that lead to growth rates driven by white-noise. The population is phenotypically structured such that each individual is associated with a value of a \textit{phenotype index}, denoted by $\varepsilon \in [0,1]$. Two individuals with the same phenotypic index are assumed to behave identically. In practice, the phenotype index may be multidimensional and determined by the expression of a potentially large number of genes; in this work, we restrict our analysis to scalar $\varepsilon$.	
	
	 In our previous work \cite{Browning.202115}, we describe a two-phenotype model of regular bacteria and persisters driven by two independent white-noise processes. We now consider a more general model through a phenotype-dependent growth rate, $\lambda(\varepsilon)$, of the form
	\begin{equation}\label{general_twonoise_growthmodel}
	\begin{aligned}
		\lambda(\varepsilon) &= \mu(\varepsilon) + s_1(\varepsilon)\xi_1 + s_2(\varepsilon)\xi_2
	\end{aligned}
	\end{equation}
	where $\xi_i$ are independent Gaussian variables (i.e., growth is driven by two independent white-noise processes) such that $\mu(\varepsilon)$ gives the time-averaged growth rate, and $s_i(\varepsilon)$ together determine the variance and correlation structure of the phenotype-dependent growth rates. One interpretation of our model is of two distinct nutrient sources, or of a nutrient and an antibiotic. The concentration of each substance is multivariate normally distributed but temporally uncorrelated; the correlation between substances allows for concentrations to, on average, increase or decrease together. The result is a set of multivariate normally distributed growth rates, with the correlation in growth rates between phenotypes forming a proxy for the relative exposure or dependence of each phenotype on each substance. 
		
		
	 Formulation of an appropriate model for the population dynamics in the presence of an arbitrary number of phenotypes is nontrivial. We follow canonical studies \cite{Kussell.2005,Acar.2008} and assume that the phenotypic state of individuals is heritable, while phenotypic transitions are distributive: all individuals switch phenotype to that chosen from a genetically-encoded \textit{target distribution}, which we denote by $\hat{p}(\varepsilon)$ (\cref{fig1}).  The switching rate, denoted by $\omega$, may potentially vary between phenotypes; for convenience we assume it to be phenotype-independent. Our premise is that populations evolve a fixed switching strategy to maximise their fitness in a given stochastically fluctuating environment. We do not study evolution, but aim to demonstrate the existence of stochastic environments in which a population that regulates some level of (discrete or continuous) phenotypic heterogeneity has a higher fitness than a homogeneous population. 

	Altogether, the concentration of cells with phenotype index $\varepsilon_i$, denoted by $r(\varepsilon_i,t)$, is governed by the stochastic partial differential equation
		\begin{equation}\label{r_spde}
			\pdv{r(\varepsilon_i,t)}{t} = \underbrace{\vphantom{\Big(}\lambda(\varepsilon_i) r(\varepsilon_i,t)}_{\text{Proliferation}} + \underbrace{\vphantom{\Big(}\omega \hat{p}(\varepsilon_i) n(t) - \omega r(\varepsilon_i,t).}_{\text{Phenotype switching}}
		\end{equation}
	 Here, $n(t) = \sum_i r(\varepsilon_i,t)$ or $n(t) = \int_0^1 r(\varepsilon,t) \,\dd \varepsilon$ for $\varepsilon$ discrete or continuous, respectively. Denoting a probability mass (or density) function for phenotypic structure of the population by $p(\varepsilon,t) = r(\varepsilon,t) / n(t)$, we write
		\begin{equation}\label{whitenoise_pop}
			\dfrac{\dd n(t)}{n} = \mathbb{E}_{p(\varepsilon,t)}\big(\mu(\varepsilon)\big)\,\dd t + \sqrt{\sum_{i=1}^2\mathbb{E}_{p(\varepsilon,t)}^2\big(s_i(\varepsilon)\big)} \; \dd W(t),
		\end{equation}
	where $W(t)$ is a Wiener process and where $\mathbb{E}_{f(\varepsilon)}$ denotes an expectation with respect to the probability mass or density function $f(\varepsilon)$. Our formulation of \cref{r_spde} is non-local: cells change phenotype according to a target distribution that does not depend on a cell's current phenotype. An extension to our model could include a diffusion or drift term to introduce local changes in phenotype throughout a continuously-structured phenotype space \cite{Chisholm.2015,DeSouzaSilva.2023}. 

	 Analysis of \cref{r_spde} to determine the population fitness as a function of the switching strategy is, in general, both analytically and numerically intractable. We make progress by instead studying fitness as a function of the \textit{expected} phenotype composition of the population, given by
		\begin{equation}
			p(\varepsilon) = \lim_{T \rightarrow \infty} \dfrac{1}{T} \int_0^T p(\varepsilon,t) \, \dd \varepsilon,
		\end{equation}
	where the fitness can now be expressed as
		\begin{equation}\label{whitenoise-fitness}
			\varphi = \mathbb{E}\left(\dv{\log n}{t}\right) = \mathbb{E}_{p(\varepsilon)}\big(\mu(\varepsilon)\big) - \dfrac{1}{2}\sum_{i=1}^2 \mathbb{E}_{p(\varepsilon)}^2 \big(s_i(\varepsilon)\big).
		\end{equation}
	While we cannot analytically determine a relationship between $\hat{p}(\varepsilon)$ and $p(\varepsilon)$, we note that should the switching strategy by homogeneous, the resultant expected composition will also be homogeneous. Likewise, a heterogeneous expected composition can only be regulated through a heterogenous switching strategy. To demonstrate scenarios where a heterogeneous switching strategy is advantageous it is, therefore, sufficient to demonstrate that a population with heterogeneous expected composition has a higher fitness than a homogeneous population. We do note, however, than $p(\varepsilon) \rightarrow \hat{p}(\varepsilon)$ in the limit that $\omega \rightarrow \infty$.

	\newpage
	\subsection{Discrete phenotypic states}\label{two_phenotype}

\begin{figure*}[!t]
			\centering
			\includegraphics[width=\textwidth]{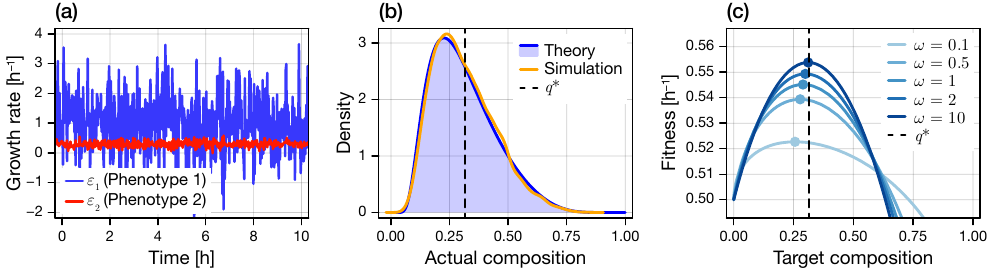}
			\caption[Fig 2]{ \textbf{Correlated white-noise growth in a discrete two-phenotype population.} (a) Correlated growth rates in a highly proliferative phenotype (blue), and a slow-growing dormant phenotype (red). The population is able to increase its effective growth rate from $\varphi(1) = \SI{0.5}{\per\hour}$ to $\varphi(q^*) \approx \SI{0.56}{\per\hour}$ using an optimal heterogeneous strategy with $p(\varepsilon_2) = q^* \approx 0.32$. (b) Stationary distribution of $x(t) = p(\varepsilon_2,t)$ (proportion of dormant cells) arising from a switching strategy with $\omega = 1$ and target distribution $\hat{p}(\varepsilon_2) = q_\text{obs}$ approximated using repeated simulation of the SDE (yellow) and theoretically through the Fokker-Planck equation (blue). (c) Numerical calculations for fitness as a function of the target distribution $\hat{p}(\varepsilon)$ for increasing values of the switching rate $\omega$. For fast switching rates, the optimal target distribution is similar to the optimal expected distribution. Even for relatively slow switching rates, the optimal target distribution is similar to the optimal controlled distribution. Parameters are set to $\mu_1 = \SI{1}{\per\hour}, \mu_2 = \SI{0.3}{\per\hour}$, $\sigma_1 = 1$, $\sigma_2 = 0.1$, $\rho = -0.5$.}
			\label{fig2}
		\end{figure*}

	 We begin our analysis by building on our previous work \cite{Browning.202115} and studying a population that can regulate only two discrete phenotypes such that $\varepsilon \in \{\varepsilon_1,\varepsilon_2\}$ \cite{Balaban:2004,Kussell.2005}. We denote by $q := p(\varepsilon_2)$ the expected proportion of the second phenotype, such that $q \in (0,1)$ corresponds to a population that regulates phenotype heterogeneity, and $q \in \{0,1\}$ corresponds to a homogeneous population. The growth rate of each phenotype is given by 
		\begin{equation}\label{two_phenotype_whitenoise}
			\lambda(\varepsilon) = \left\{\begin{array}{ll}
				\mu_1 + \sigma_1 \xi_1, & \varepsilon = \varepsilon_1,\\
				\mu_2 + \sigma_2 \rho \xi_1 + \sigma_2 \sqrt{1 - \rho^2} \xi_2, & \varepsilon = \varepsilon_2.
		\end{array}\right.
		\end{equation}
	Both subpopulations experience normally distributed, temporally uncorrelated growth rates. The mean and variance of the growth rate experienced by the subpopulation with phenotype $\varepsilon = \varepsilon_1$ is given by $\mu_i$ and $\sigma_i^2$.  The correlation between $\lambda(\varepsilon_1)$ and $\lambda(\varepsilon_2$) is determined by the parameter $\rho$. An example set of temporally fluctuating growth rates is shown in \cref{fig2}a. 
	
	 The effective growth rate of each phenotype, and therefore the fitness in an isolated homogeneous population,  is given by
		\begin{equation}
			\varphi_i = \mu_i - \frac{\sigma_i^2}{2}.
		\end{equation}
	 Without loss of generality, we assume that the phenotypes are ordered such that $\varphi_2 < \varphi_1$.  For bet-hedging systems (e.g., bacterial persistence), we may additionally assume that $\mu_2 \ll 1$, however, this need not be the case.  For example, different phenotypes may represent a specialisation to different nutrient sources: the expected growth rates may be similar, $\mu_1 \approx \mu_2$, with $\rho < 0$ such that the concentration of one nutrient corresponds to a likely decrease in the other. 
		
	 Substituting the growth rate model (\cref{two_phenotype_whitenoise}) into \cref{whitenoise-fitness} yields a quadratic expression for the fitness as a function of $q \in [0,1]$, denoted by $\varphi(q)$ (a full derivation is provided in the supplementary material). Employing heterogeneity will, therefore, be advantageous should $q^* = \mathrm{arg\,max}\,\varphi(q)$ be on the interior $q^* \in (0,1)$. Given that $\varphi(0) = \varphi_1 > \varphi_2 = \varphi(1)$ by our previous assumption, this will be the case provided that (1) $\varphi''(q)$ is strictly negative, and that (2) $q^* > 0$.  Both conditions hold provided that
		\begin{equation}\label{whitenoise_cond1}
			\mu_2 > \mu_1 + \rho \sigma_1 \sigma_2 - \sigma_1^2,
		\end{equation}
	in which case the optimal proportion is given by
		\begin{equation}\label{whitenoise_qstar}
			q^* 	= \dfrac{\mu_2 - \mu_1 + \sigma_1^2 - \rho\sigma_1\sigma_2}{\sigma_1^2 - 2\rho\sigma_1\sigma_2 + \sigma_2^2}.
		\end{equation}
	
	Given that $\varphi_1 > \varphi_2$, which places an upper bound on $\mu_2$, we further note that heterogeneity is only advantageous in the white-noise environment model if
		\begin{equation}\label{whitenoise_cond2}
			\dfrac{\sigma_1^2 + \sigma_2^2}{2}  > \rho\sigma_1 \sigma_2.
		\end{equation}
	Clearly, the range of scenarios in which a two-phenotype strategy is advantageous is reduced as $\rho \rightarrow 1$ and increases as $\rho \rightarrow -1$. For the example presented in \cref{fig2}a, the population may employ phenotypic heterogeneity to increase its fitness by approximately 12\% from \SI{0.5}{\per\hour} (with $p(\varepsilon_2) = 0$) to \SI{0.56}{\per\hour} (with $p(\varepsilon_2) = q^* \approx 0.32)$. Other parameter values are given in the caption of \cref{fig2}.
	
	\subsubsection{Finite phenotypic switching rate}

	To investigate the effect the switching rate $\omega$ on fitness, we consider that the actual phenotype composition $p(\varepsilon_2,t) := x(t)$ is itself governed by an SDE (full details are provided in the supplementary material). The stationary distribution for $x(t)$ is given by a boundary value problem on $x(t) \in [0,1]$ that arises out of the one-dimensional Fokker-Planck equation. This equation admits, up to a normalisation constant, an analytical solution that can be used to numerically compute the expectations required to determine the fitness for general $\omega$ (\cref{whitenoise-fitness}). We take the population to target the optimal proportion of the second phenotype such that $\hat{p}(\varepsilon_2) = q^*$. Results in \cref{fig2}b show that the actual composition, $x(t)$, fluctuates across a wide range of values, and that our theoretical solution for the distribution of $x(t)$ is in agreement with simulation results. 
	
	Results in \cref{fig2}c show the fitness as a function of $\omega$ and the now target dormant proportion $q_\mathrm{opt}$, in addition to the optimal target composition for each $\omega$. First, we see that it is indeed true that $p(\varepsilon)$ converges to $\hat{p}(\varepsilon)$ for sufficiently large $\omega$. Secondly, we find that the optimum remains relatively close to the large $\omega$ limit, although the fitness advantage of the homogeneous strategy is reduced for small $\omega$. For white-noise environments, we conjecture that the fitness may be a monotonic function of the switching rate: specifically, a monotonically increasing function in cases where heterogeneity is advantageous, and a monotonically decreasing function in cases where the homogeneous strategy is favoured.

	
	\subsection{Continuously distributed phenotypes}
	
	\begin{figure}[!t]
		\centering
		\includegraphics[width=\textwidth]{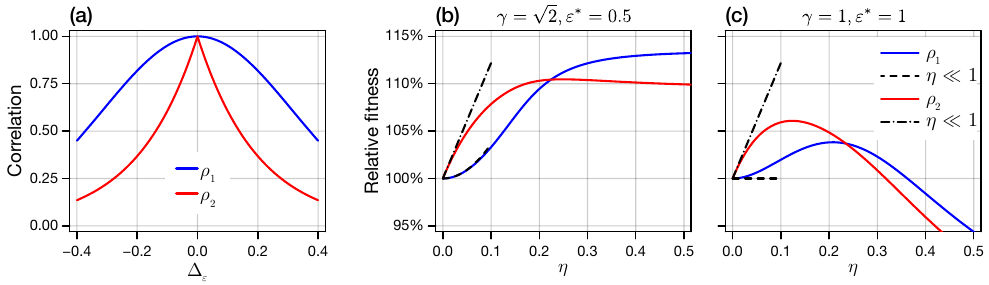}
		\caption[Fig 3]{\textbf{Unimodal model of continuous phenotypic variability.} (a) We consider correlation functions that are either continuously differentiable at $\Delta_\varepsilon = 0$ (blue; $\rho_1(\Delta_\varepsilon) = \mathrm{exp}(-5\Delta_\varepsilon^2)$) or not (red; $\rho_2(\Delta_\varepsilon) = \mathrm{exp}(-5|\Delta_\varepsilon|)$). (b--c) The fitness relative to $\eta = 0$ is computed as a function of phenotype variance $\eta^2$ using numerical quadrature (coloured curves) and the asymptotic expressions (black). The optimal phenotype $\varepsilon^*$ is (b) on the interior of the domain at $\varepsilon^* = 0.5$ and (c) at $\varepsilon^* = 1$. In (b--c), the variance model is given by \cref{quadratic_variance} with $k = 1$ and (b) $\gamma = \sqrt{2}$ and (c) $\gamma = 1$.}
		\label{fig3}
	\end{figure}
	
	We now consider that an otherwise homogeneous population undergoes mutations that yield a continuous distribution of phenotypes. We assume that the distribution is initially unimodal, and develops over relatively slow evolutionary timescales so that the initial variance of the expected phenotype distribution, denoted by $\eta^2$, is sufficiently small.  Such variability in phenotype could be, e.g., regulated through variability in stochastic gene expression.  Our goal is to determine whether or not such small amounts of subpopulation variability yield a fitness advantage over a homogeneous population.	 In \cref{contbimodal}, we extend our analysis to a bimodal model where two discrete phenotypes develop subpopulation heterogeneity simultaneously.

		\subsubsection{Unimodal strategy}

		 The first model of continuous phenotypic variability is unimodal. An initially homogeneous and evolutionarily stable population undergoes small mutations that lead to relatively small differences in growth rate. We model this by assuming that $\varepsilon \sim \mathcal{N}_{[0,1]}(\varepsilon^*,\eta^2)$ where $\varepsilon^*$ denotes the optimal single phenotype, and where $\mathcal{N}_{[0,1]}$ denotes a truncated Gaussian distribution on $\varepsilon \in [0,1]$. In white-noise environments, the growth rate of each phenotype is given by
			\begin{equation}
				\lambda(\varepsilon) := \mu(\varepsilon) + \sigma(\varepsilon) \xi
			\end{equation}
		for some $\xi \sim \mathcal{N}(0,1)$. We also assume that both $\mu(\varepsilon)$ and $\sigma(\varepsilon)$ are monotonically increasing functions of $\varepsilon$. 
		
		To provide a set of analytically tractable and interpretable results, we first focus our analysis on the case that $\sigma^2(\varepsilon)$ is quadratic with form
			\begin{equation}\label{quadratic_variance}
				\sigma^2(\varepsilon) = \gamma^2\big[k (\varepsilon^2 - \varepsilon) + \varepsilon \big],
			\end{equation}
		such that $k$ represents the curvature and $\gamma^2$ represents the variance at $\varepsilon = 1$. For $k = 0$, the variance scales linearly with growth rate, for $k < 0$ the variance is convex such that the variance increases more quickly at smaller values of $\varepsilon$, and for $k > 0$ a faster increase in variance is seen at larger values of $\varepsilon$. We additionally consider that $\mu(\varepsilon) = \varepsilon$ with a corresponding maximum at  $\mu(1) = \SI{1}{\per\hour}$ . This latter choice can be made without loss of generality as the phenotype index itself is arbitrary and a temporal rescaling can always be applied to enforce a given maximum growth rate.
		
		 We first determine the evolutionarily stable homogeneous strategy by finding the single phenotype, $\varepsilon^*$, that maximises population fitness.  It can easily be seen that phenotypes on the interior $\varepsilon^* \in (0,1)$ can only be optimal for $k > 0$. More generally, this observation is true for all concave variance functions: a necessary (but not sufficient) condition for the optimal phenotype to lie on the interior of the domain is that $\sigma^2(\varepsilon)$ is \textit{not} linear or strictly concave. For the quadratic model, we state the optimal phenotype for various $\gamma$ in \cref{tab1}.
	
	\begin{table}[!b]
		\centering
		\caption{Optimal single phenotype $\varepsilon^*$ for the continuously distributed phenotype model with quadratic variance function given by \cref{quadratic_variance}. The parameter $\gamma^2$ determines the growth rate variance at $\varepsilon = 1$, and $k$ determines the concavity (or convexity) of the variance as a function of $\varepsilon$.}
		\label{tab1}
		\renewcommand{\arraystretch}{1.3}
		\begin{tabular}{|c|c|c|}\hline
			& $\gamma^2 < 2$ & $\gamma^2 > 2$\\\hline\hline
			Linear & 1 & 0\\\hline
			Convex & \multicolumn{2}{c|}{$\min\left(1,\max\left[0,\frac{1}{2} + \frac{1}{k}\left(\frac{1}{\gamma^2} - \frac{1}{2}\right)\right]\right)$} \\\hline
			Concave & 1 & 0 \\\hline
		\end{tabular}
	\end{table}
		
		To extend the growth model to a continuum of phenotypes, we define a correlation function,
		\begin{equation}
			\rho(\Delta_\varepsilon) = \mathrm{Corr}\big(\lambda(\varepsilon^* + \Delta_\varepsilon),\lambda(\varepsilon^*)\big),
		\end{equation}
		that describes the correlation between growth rates, relative to the optimal phenotype, $\varepsilon^*$. We always assume that the correlation function is an even function. This formulation corresponds to our general model (\cref{general_twonoise_growthmodel}) with
			\begin{subequations}
			\begin{align}
				s_1(\varepsilon) &= \rho(\varepsilon - \varepsilon^*)\sigma(\varepsilon),\\
				s_2(\varepsilon) &= \sigma(\varepsilon)\sqrt{1 - \rho^2(\varepsilon - \varepsilon^*)}.
			\end{align}
			\end{subequations}

		The integrals required to compute the expected growth rate are, in general, intractable. We make progress by considering that $0 < \eta \ll 1$ and studying the behaviour of $s_1(\varepsilon)$ and $s_2(\varepsilon)$ within a small neighbourhood of $\varepsilon^*$.  Consequentially, the results that follow apply to general $\sigma(\varepsilon)$.  The sign of the first-order correction term thus determines whether continuous heterogeneity is advantageous for small $\eta$, and can be interpreted as a sufficient (albeit not necessary) condition for heterogeneity to present some advantage.

		We first consider cases where the optimal single phenotype, $\varepsilon^*$, lies on the interior of the domain. Results can be summarised by two subcases, which depend on the behaviour of the correlation function near $\Delta_\varepsilon = 0$ (as shown in \cref{fig3}a):
		\begin{enumerate}
			\item \textit{$\rho(\Delta_\varepsilon)$ is differentiable at $\Delta_\varepsilon = 0$.} To leading order, we have that 
				\begin{equation}\label{whitenoise_cont_1}
					\varphi_\eta = \varphi_0 - \underbrace{\vphantom{\bigg(}
					\dfrac{\sigma(\varepsilon^*)\Big((\pi - 2)\sigma(\varepsilon^*)\rho''(0) + \pi \sigma''(\varepsilon^*)\Big)}{2\pi}
					}_{\text{Leading order correction}}\eta^2 + \mathcal{O}(\eta^3).
				\end{equation}
				
			\item \textit{$\rho(\Delta_\varepsilon)$ is not differentiable at $\Delta_\varepsilon = 0$}. To leading order, we have that
				\begin{equation}
				\begin{aligned}
					\varphi_\eta &= \varphi_0 + \dfrac{\sqrt{2}\sigma^2(\varepsilon^*) \left|\rho'_-(\varepsilon^*)\right|\left(\sqrt{\pi} - \Gamma^2\left(\tfrac{3}{4}\right)\right)}{\pi}\eta + \mathcal{O}(\eta^2),
				\end{aligned}
				\end{equation}
				where $\rho'_-(\varepsilon^*) := \lim_{\Delta_\varepsilon \rightarrow 0^-} \rho'(\Delta_\varepsilon)$,  and where $\Gamma(z)$ is the gamma function. 
		\end{enumerate}
		In the second case we can readily see that the leading order correction term is positive, and conclude therefore that there exists (at least sufficiently small) choices of $\eta$ such that $\varphi_\eta \ge \varphi_0$, for all choices of $\sigma(\varepsilon)$. In the first case, heterogeneity will be advantageous provided that
		\begin{equation*}
			|\rho''(0)| > \dfrac{\pi \sigma''(\varepsilon^*)}{(\pi - 2) \sigma(\varepsilon^*)},
		\end{equation*}
		a condition that will always hold for concave $\sigma(\varepsilon)$, and will also hold for convex $\sigma(\varepsilon)$ provided that $|\rho''(0)|$ is sufficiently large. We verify these results in \cref{fig3}b by comparing the asymptotic expansion for the fitness to that calculated numerically through quadrature.
		
		We proceed with a similar analysis for the case where $\varepsilon^* = 1$ on the exterior of the domain. The approach is similar, however asymptotically $\varepsilon$ now has a half-Gaussian distribution with scale parameter $\eta$ that can no longer be interpreted directly as the phenotype variance (the variance in this case is given by $(1 - 2/\pi)\eta^2$). Similarly, we can characterise the results into two cases that depend on the behaviour of the correlation function at $\Delta_\varepsilon = 0$: 
		\begin{enumerate}
			\item \textit{$\rho(\Delta_\varepsilon)$ is differentiable at $\Delta_\varepsilon = 0$.} To leading order, we have that 
				\begin{equation}
					\varphi_\eta = \varphi_0 + \sqrt{\dfrac{2}{\pi}} \left(\sigma(1)\sigma'(1) - 1\right) \eta + \mathcal{O}(\eta^2).
				\end{equation}
				
			\item \textit{$\rho(\Delta_\varepsilon)$ is not differentiable at $\Delta_\varepsilon = 0$}. To leading order, we have that
				\begin{equation}\label{whitenoise_cont_4}
				\begin{aligned}
					\varphi_\eta \sim \varphi_0 &+ \sqrt{\dfrac{2}{\pi}}\Bigg(\sigma(1) \Big(\sigma(1)\rho'_-(0) + \sigma'(1)\Big) -\dfrac{\sigma^2(1)\rho'_-(0)\Gamma(\frac{3}{4})^2}{\sqrt{\pi}} - 1\Bigg)\eta + \mathcal{O}(\eta^2).
				\end{aligned}
				\end{equation}
		\end{enumerate}
		In the first case, heterogeneity will be advantageous if $\sigma(1)\sigma'(1) > 1$. For the quadratic model, this will hold provided $(1 + k)\gamma^2 > 2$. We again verify these results in \cref{fig2}c by comparing the asymptotic expansion for the fitness to that calculated numerically through quadrature.

		\subsubsection{Bimodal strategy}\label{contbimodal}
		
		\begin{figure}[!t]
			\centering
			\includegraphics[]{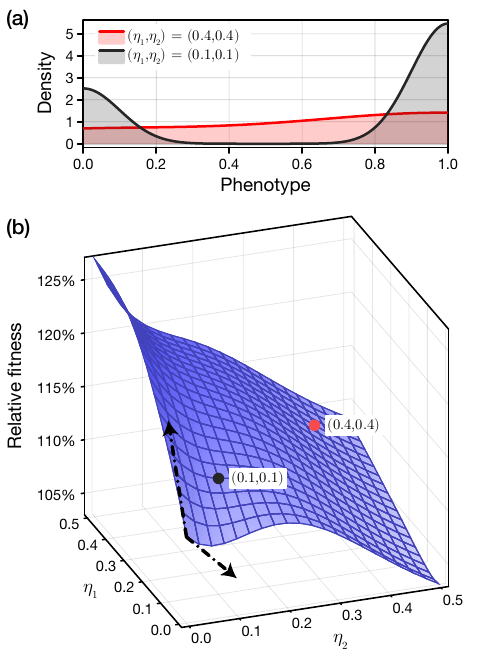}
			\caption[Fig 4]{\textbf{Bivariate continuous model of heterogeneity.} We consider that the population in \cref{fig2}a develops continuous heterogeneity at both $\varepsilon = 0$ (with variance $\eta_1^2$) and at $\varepsilon = 1$ (with variance $\eta_2^2$), such that the target distribution follows a mixture of truncated Gaussian distributions. (a) Example target distributions for $(\eta_1, \eta_2) = (0.4,0.4)$ and $(\eta_1, \eta_2) = (0.1,0.1)$. For either $\eta_1, \eta_2 \gg 0$, the target distribution approaches a uniform distribution (this is seen clearly for $\eta_1 = \eta_2 = 0.4$). (b) Numerical approximation of the fitness, relative to that for a homogeneous strategy with all cells at $\varepsilon = 1$. Shown also is the perturbation expansion for $\eta_1, \eta_2 \ll 0$ (\cref{binary_perturb}).}
			\label{fig4}
		\end{figure}
		
		The next model of continuous heterogeneity that we consider is based on the  two-phenotype model  studied in \cref{two_phenotype}. We assume that a population develops continuous heterogeneity from an otherwise evolutionary stable two-phenotype strategy, such that the expected phenotype composition is a mixture
			\begin{equation}\label{discrete_mixture}
				p(\varepsilon) = q^*\mathcal{N}_{[0,1]}(0,\eta_1) + (1 - q^*)\mathcal{N}_{[0,1]}(1,\eta_2),
			\end{equation}
		where $q^*$ is the optimal proportion of individuals at $\varepsilon = 0$ (\cref{whitenoise_qstar}).
		
		For correlation functions that are differentiable at both $\Delta_\varepsilon = 0$ and $\Delta_\varepsilon = 1$, the fitness is given, to leading order, by
			\begin{equation}\label{binary_perturb}
				\varphi_{(\eta_1,\eta_2)} = \varphi_0 + \sqrt{\dfrac{2}{\pi}} \Big(c_1 q^* \eta_1 + c_2 (1 - q^*)\eta_2\Big),
			\end{equation}
		where
			\begin{align*}
				c_1 &= \mu'(0) - (1 - q^*) \sigma'(0) \sigma(1) \rho(1) - \Big(q^* \sigma'(0) + (1 - q^*) |\rho'(1)| \sigma(1)\Big) \sigma(0),\\
				c_2 &= -\mu'(1) - q^* \sigma(0)\sigma(1)\sqrt{|\rho''(0)|(1 - \rho(1)^2)} + \Big(q^* \sigma(0) \rho(1) + (1 - q^*)\sigma(1)\Big)\sigma'(1).
			\end{align*}
		Thus, continuous heterogeneity is advantageous if either $c_1$ or $c_2$ are positive. 
		
		In \cref{fig4}, we explore the effect of continuous heterogeneity arising out of the discrete strategy demonstrated in \cref{fig2}. Here, we take $\sigma(\varepsilon)$ to be of a quadratic form with $k = 0.3$ and $\rho(\Delta_\varepsilon)$ to be differentiable, with all other parameters chosen such that $\sigma(0) = \sigma_2$, $\sigma(1) = \sigma_1$, and $\rho(1) = \rho$ correspond to earlier analysis in \cref{fig2}. In \cref{fig4}a, we demonstrate that \cref{discrete_mixture} tends to a uniform distribution for $\eta_1, \eta_2 \gg 1$. Results \cref{fig4}b show, both numerically and through the perturbation expansion, that continuous heterogeneity presents a fitness advantage in the neighbourhood of $\varepsilon = 0$, and a fitness disadvantage near $\varepsilon = 1$. 
		
\FloatBarrier

\section{Growth in temporally  correlated environments }\label{sec_continuous}

\begin{figure}[!t]
		\centering
		\includegraphics[width=\textwidth]{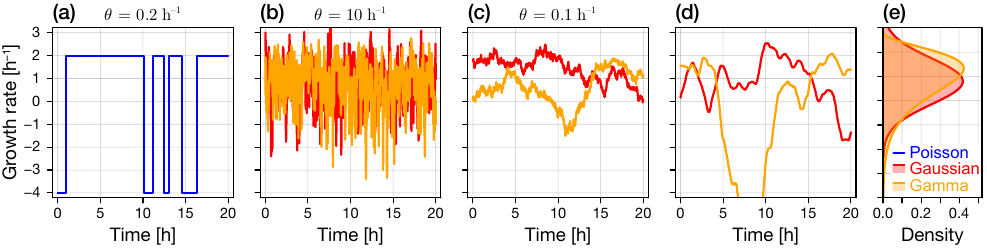}
		\caption[Fig 5]{ \textbf{Growth rate of the fittest phenotype in temporally correlated environments.} (a) A homogeneous Poisson process drives transitions between a ``growth'' state, characterised by positive net growth rates, and a ``stress'' state, characterised by negative net growth rates. (b--d) Fluctuating environments with stationary growth rate distributions, shown in (e), that are either Gaussian (red) or Gamma (orange) distributed. Shown in (b,c) are growth rates driven by Ornstein-Uhlenbeck processes that are relatively fast (b) or slow (c). In (d), the growth rate is driven by a second-order Ornstein-Uhlenbeck process such that the derivative of the growth rate is also continuously differentiable. All environments have an identical mean growth rate, and identical mean time spent with a negative growth rate.}
		\label{fig5}
	\end{figure}

	 We now turn to phenotypic heterogeneity in temporally correlated environments. Specifically, we study environments that lead to growth rates driven by homogeneous Poisson processes, and growth rates in which \textit{changes} are driven by white-noise processes and are, therefore, differentiable everywhere. An example suite of such environments is given in \cref{fig5}. 
	
	 The cellular dynamics are identical to before, and are described by
		\begin{equation}\label{r_spde2}
			\pdv{r(\varepsilon,t)}{t} = \underbrace{\vphantom{\Big(}\lambda(\varepsilon,z(t)) r(\varepsilon,t)}_{\text{Proliferation}} + \underbrace{\vphantom{\Big(}\omega \,\hat p(\varepsilon) n(t) - \omega \,r(\varepsilon,t).}_{\text{Phenotype switching}}
		\end{equation}
	Compared to \cref{r_spde}, we are now explicit in our relation of the phenotype-dependent growth rate, denoted by $\lambda(\varepsilon,z(t))$, to the environment, denoted by $z(t)$. In this work, we restrict our analysis to environmental fluctuations that are, in some sense, one-dimensional  (e.g., driven by fluctuations in a single nutrient or antibiotic \cite{Kussell.2005,Acar.2008}, or by fluctuations in temperature \cite{Charlebois.2018,Nguyen.2021}).  We again assume that the target distribution, $\hat{p}(\varepsilon)$, is genetically-encoded and, therefore, fixed. Implicit in this assumption for temporally correlated environments is that the timescale of environmental fluctuations is not so slow that the population adapts the target distribution to the current state of the environment through evolution. 
				
	 It is useful to rewrite \cref{r_spde2} in terms of the total cell concentration and the phenotype distribution, denoted by $n(t)$ and $p(\varepsilon,t) = r(\varepsilon,t) / n(t)$, respectively. 	Appropriate substitutions yield the coupled equations
		\begin{subequations}
		\begin{align}
			\dv{\log n(t)}{t} &= \int_0^1 \lambda(\varepsilon,z(t)) p(\varepsilon,t) \dd \varepsilon,\label{cont_logn}\\
			\pdv{p(\varepsilon,t)}{t} &= p(\varepsilon,t)\Big[\lambda(\varepsilon,z) - \mathbb{E}_{p(\varepsilon,t)} \big(\lambda(\varepsilon,z)\big) - \omega\Big] + \omega \hat{p}(\varepsilon)\label{cont_p},
		\end{align}
		\end{subequations}
	where we see that the dynamics of $p(\varepsilon,t)$ do not depend on $n(t)$. From \cref{cont_logn}, we obtain an expression for the fitness
		\begin{equation}\label{cont_fitness}
			\varphi = \mathbb{E}_W\mathbb{E}_{p(\varepsilon,t)} \big(\lambda(\varepsilon,z)\big),
		\end{equation}
	where the outer expectation is taken with respect to the measure, denoted by $W$, driving fluctuations in the environment, $z(t)$. This expression highlights the primary difficulty in moving from white-noise environments to temporally correlated environments: in general, we now expect the composition of the population, through $p(\varepsilon,t)$, to be correlated with the environment $z(t)$, which is never the case for environments that are temporally uncorrelated. For example, when environmental conditions permit growth, we expect the population to move towards a composition dominated by growing phenotypes. Conversely, when the growth rate is low (or negative), we expect dormant phenotypes to become dominant.  In the limit that environmental fluctuations are slow compared to the timescales relating to phenotype switching, we expect this correlation to become perfect such that $p(\varepsilon,t) = \tilde{p}(\varepsilon,z)$ for some function $\tilde{p}$. Conversely, in the limit that environmental fluctuations are faster than that of phenotype switching (denoted the \textit{fast regime}), we expect $p(\varepsilon,t)$ to become independent of $z(t)$ and so the fitness is determined solely by the time-averaged growth rate for each phenotype. 
	
	 \Cref{cont_fitness} is, in general, both analytically and computationally intractable. Even for a discrete model in which $p(\varepsilon,t)$ comprises $n$ discrete phenotypes, numerical approaches to calculate the fitness will involve the solution of an $n$-dimensional Fokker-Planck equation for the joint stationary density of $z(t)$ and $p(\varepsilon,t)$. Our goal, however, is not to compute the fitness of a population in a given environment, but to present a framework for determining the existence of environments in which phenotypic heterogeneity presents a fitness advantage.  Where necessary, we make progress by studying the \textit{slow regime}, in which environmental fluctuations occur on a significantly slower timescale to the cellular dynamics. In the slow regime, we can apply a \textit{quasi-steady state} (QSS) approximation to \cref{cont_p}, denoted $\tilde{p}(\varepsilon,z)$, where, on the cellular dynamics timescale (the fast timescale), the current state of the $z$ is assumed to be fixed. The QSS approximation is given as the solution to the algebraic equation
		\begin{equation}\label{qss1}
			0 = \tilde{p}(\varepsilon,z)\Big[\lambda(\varepsilon,z) - \mathbb{E}_{\tilde{p}(\varepsilon,z)} \big(\lambda(\varepsilon,z)\big) - \omega\Big] + \omega \hat{p}(\varepsilon),
		\end{equation}
	which is a root finding problem in $\mathbb{E}_{\tilde{p}(\varepsilon,z)}\big(\lambda(\varepsilon,z)\big)$, the primary quantity of interest in the calculation of the fitness (\cref{cont_fitness}). We note that \cref{qss1} is only well-posed for $\lambda(\varepsilon,z) - \omega \neq \mathbb{E}_{\tilde{p}(\varepsilon,z)}\big(\lambda(\varepsilon,z)\big)$ for all $\varepsilon$. If this condition does not hold, a degeneracy will form in the solution to $\tilde{p}(\varepsilon,z)$, corresponding to domination by a single phenotype  (i.e., a homogeneous population). 	
	
	
	\subsection{Poisson model of discrete transitions}
	
	The most common theoretical and experimental model of environmental fluctuations comprises discrete transitions between a finite number of states (e.g., presence or absence of antibiotic \cite{Kussell.2005}, or transitions between nutrient sources \cite{Lambert.2014}). The Poisson model \cite{Wolf.2005,Muller.201303s} represents a stochastic analogue of the periodically switching environments studied by Kussell et al. \cite{Kussell.2005}. Here, the environment may take one of a finite number of discrete states, where switches between states are governed by an inhomogeneous Poisson process. In \cref{fig5}a, we demonstrate an environment that switches between a ``growth state'' and a ``stress state''.  The net growth rate of the fittest phenotype ($\varepsilon = 1$) in each state is chosen to match that presented in \cite{Kussell.2005}.

	In the Poisson model, state transitions are fully characterised by the expected time spent in each state, denoted by $\tau_0 := \alpha \theta^{-1} $ and $\tau_1 := (1 - \alpha) \theta^{-1}$ for the stress and growth states, respectively. Here, the parameter $\alpha$ corresponds to the proportion of time spent in the stress state, and $\theta^{-1}$ determines the timescale: $\theta \rightarrow 0$ corresponds to the slow regime, and $\theta \rightarrow \infty$ to the fast regime. The growth rates of each phenotype depend only on the current state of the environment. In this section, we follow \cite{Kussell.2005v2} and consider only a two-phenotype model---regular and dormant---and assume for simplicity that dormant cells ($\varepsilon = 0$) are completely dormant, with a constant net growth rate of zero.
	
\begin{figure}[!t]
		\centering
		\includegraphics[]{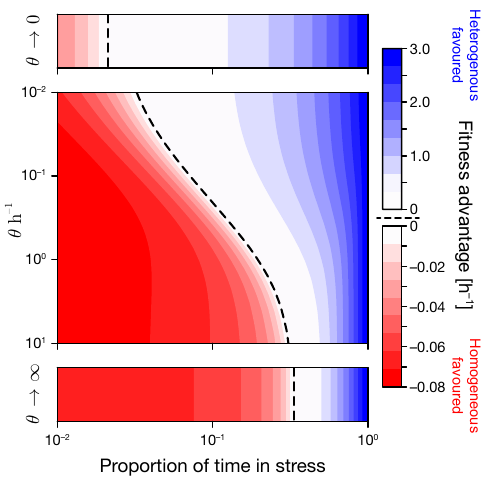}
		\caption[Fig 6]{\textbf{Discrete heterogeneity in an environment with fluctuations driven by a Poisson process.} Phenotype switching occurs according to a target distribution with $q = 0.1$ at rate $\omega = \SI{1}{\per\hour}$. The environment switches between growth and stress conditions (see \cref{fig5}a) for a given proportion of time in stress such that the average cycle has duration $\theta^{-1}$. The fitness advantage, relative to that of a homogeneous strategy with $q = 0$, is calculated using a numerical solution to the Fokker-Planck equation for general $\theta$. Note that regions of advantage and disadvantage are shown on a different scale. We also calculate the QSS approximation in the slow regime limit ($\theta \rightarrow 0$) and the fast limit ($\theta \rightarrow \infty$).}
		\label{fig6}
	\end{figure}

	In \cref{fig6}, we compare a heterogeneous strategy, where cells target small proportion, $\hat{p}(0) = 0.1$, of cells at the dormant phenotype with switching rate $\omega = \SI{1}{\per\hour}$, to a homogeneous strategy across a range of environmental parameters. Results for general $\theta$ are constructed from an approximate numerical solution to the Fokker-Planck equation for the joint distribution of the proportion of dormant cells, $x(t)$, and the environment, $z(t)$ (see the supplementary material). The fitness in both the fast and slow regimes is available analytically. In the limit that the environment changes quickly (i.e., $\theta \rightarrow \infty$), the heterogeneous strategy is only advantageous in very stressful environments (i.e., for $\alpha \rightarrow 1$), whereas in the limit that the environment changes slowly (i.e., $\theta \rightarrow 0$), the heterogeneous strategy is favoured for a range of $\alpha$. Importantly, we see that the potential fitness advantage gained through a heterogeneous strategy is markedly high in comparison to the magnitude of any potential advantage gained through a homogeneous strategy. 
	
	\subsection{Continuously distributed growth rates}
	
	 We now extend our framework to a more general model of the environment, characterised by the long-term distribution of growth rates for each phenotype, denoted by $\mathcal{D}(\varepsilon)$. We consider that the environment, $z(t)$, is now governed by a scalar mean-reverting Ornstein-Uhlenbeck process
		\begin{equation}\label{zou}
			\dd z(t) = -\theta z(t) \dd t + \sqrt{2\theta} \,\dd W_t,
		\end{equation}
	with stationary distribution $z(t) \sim \mathcal{N}(0,1)$. The reversion strength parameter $\theta$ carries an analogous interpretation to that in the Poisson process model and determines the relative timescale of environmental fluctuations. To relate the environment to the growth rate of each phenotype we apply the probability inverse transformation such that
		\begin{equation}\label{zou_transform}
			\lambda(\varepsilon,z) = F^{-1}_{\mathcal{D}(\varepsilon)}\big(\Phi(z)\big),
		\end{equation}
	where $\Phi(\cdot)$ is the distribution function for the standard Gaussian distribution, and $F^{-1}_{\mathcal{D}(\varepsilon)}(\cdot)$ is the quantile function for $\mathcal{D}(\varepsilon)$. For simplicity, we specify only the growth rate distribution for the phenotype $\varepsilon = 1$, $\mathcal{D}(1)$, and assume that general $\mathcal{D}(\varepsilon)$ is determined through
		\begin{equation}\label{zou_transform_simple}
			\lambda(\varepsilon,z) = \varepsilon \lambda(1,z) = \varepsilon F^{-1}_{\mathcal{D}(1)}\big(\Phi(z)\big).
		\end{equation}
	That is, the phenotype index corresponds simply to an overall scaling of the growth rate and is, therefore, a proxy for how sensitive each phenotype is to the environment: $\varepsilon = 0$ corresponds to complete dormancy, $\varepsilon = 1$ to complete sensitivity. Under the assumption that the growth rates remain ordered within phenotype space, our choice of a linear scaling is arbitrary since $\varepsilon$ may be itself rescaled without loss of generality. Finally, we note that the fitness in a homogeneous strategy is given simply by the expected value of $\mathcal{D}(1)$.  
		
	In \cref{fig5}b,c, we simulate a pair of environments in both the fast ($\theta = 10$) and slow ($\theta = 0.1$) regimes. Distinguishing the environments is the long-term distribution of growth rates; we consider both a symmetric Gaussian distribution and a translated Gamma distribution with negative skewness (\cref{fig5}e). The mean growth rate and the proportion of time spent with a negative growth rate (i.e., in stress) are chosen to match that of the Poisson environment shown in \cref{fig5}a and studied in the previous section.  In \cref{fig5}d we show realisations of a corresponding pair of environments driven by a second-order Ornstein-Uhlenbeck process, such that the growth rate is not only continuous but also smooth.

		
		\subsubsection{Two discrete phenotypic states}
	
		We begin our analysis by again considering a population  that can regulate only two discrete phenotypes, $\varepsilon \in \{0,1\}$.  In the two-phenotype model, phenotype switching can be expressed by considering the switching rates directly, denoted by $a$ and $b$, such that
			\begin{equation}\label{r0r1}
			\begin{aligned}
				\dv{r_0(t)}{t} &= \lambda(0,z) r_0(t) - a r_0(t) + b r_1(t),\\
				\dv{r_1(t)}{t} &= \lambda(1,z) r_1(t) + a r_0(t) - b r_1(t),\\
			\end{aligned}
			\end{equation}
		where $r_\varepsilon(t)$ denotes the density of individuals in phenotypic state $\varepsilon$. The rates $a$ and $b$ are related to the target distribution and switching rate through $a = \omega (1 - q)$ and $b = \omega q$ where $q := \hat{p}(0)$. We again denote by $p(0,t) = x(t)$ such that $p(1,t) = 1 - x(t)$, where $x(t)$ is the proportion of cells in the dormant state. It follows from \cref{r0r1} that the dynamics of $x(t)$ are governed by
			\begin{equation}\label{dxdt}
				\dv{x(t)}{t} = h(x,z) := b - x(t) \Big(a + b + \big(\lambda(1,z) - \lambda(0,z)\big) \big(1 - x(t)\big) \Big),
			\end{equation}
		which is coupled to the SDE governing the environment (\cref{zou}), and where $\lambda(\varepsilon,z)$ is given by \cref{zou_transform_simple}.  
	
		For general $\theta$, the fitness can be expressed as the integral
			\begin{equation}\label{fitness_gentheta}
				\varphi = \int_{-\infty}^\infty	\int_0^1 \big(x \lambda(0,z) + (1 - x)\lambda(1,z) \big) \,u(x,z) \,\dd x \,\dd z,
			\end{equation}
		where $u(x,z)$ is the joint stationary distribution governed by a two-dimensional stationary Fokker-Planck equation which may be solved numerically; difficulties associated with the lack of a diffusion term in \cref{dxdt} can be overcome by introducing a small diffusion term with coefficient $\beta = 10^{-5}$ (full details are available in the supplementary material).  This inclusion provides good agreement with simulation results for large $\theta$, however produces poor results in the slow regime as $\theta \rightarrow 0$. Within the slow regime, we apply the QSS approximation to \cref{dxdt}, taking $x(t)$ to be at its steady state yielding a functional relationship between $x(t)$ and $z(t)$ which we denote $x = \tilde{x}(z)$.  Applying the law of iterated expansion, the fitness in the slow regime is given by 
			\begin{equation}\label{fitness_qss}
				\varphi_\text{QSS} = \int_{-\infty}^\infty \big(\tilde{x}(z) \lambda(0,z) + (1 - \tilde{x}(z))\lambda(1,z) \big) \Phi'(z) \,\dd z,
			\end{equation}
		where $\Phi'(z)$ is the density function for the standard Gaussian distribution.
		
\begin{figure}[!t]
			\centering
			\includegraphics[width=\textwidth]{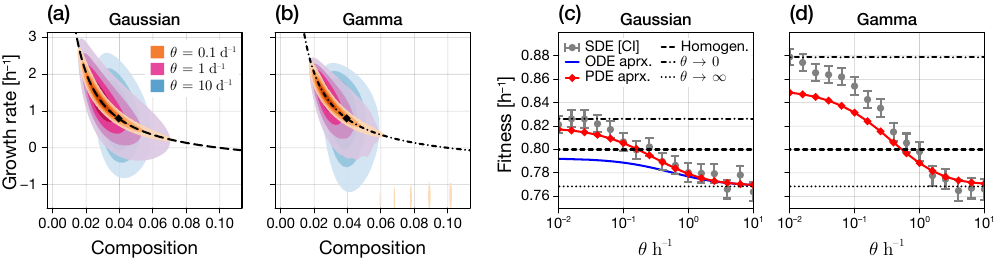}
			\caption[Fig 7]{\textbf{Discrete heterogeneity in continuously fluctuating environments.} (a,b) An approximate solution to the Fokker-Planck equation for the joint distribution of the $\varepsilon = 1$ growth rate and proportion of cells in the dormant state (for a target distribution with $q = 0.1$ with switching rate $\omega = \SI{0.5}{\per\hour}$). Results shown for the (a) Gaussian and (b) Gamma environments. (c,d) We approximate the fitness as a function of $\theta$ through repeated simulation of the SDE (50,000 realisations, shown is the mean and a 95\% confidence interval). We approximate the fitness using (red) a numerical solution to the  Fokker-Planck  PDE and (blue, for the Gaussian environment only) a ODE-based moment approximation to the SDE. Both approximations are valid for $\theta$ sufficiently large. Also shown is the QSS fitness (dash-dot, $\theta \rightarrow 0$); fast fitness (dot, $\theta \rightarrow \infty$); and the fitness for the homogeneous strategy (black dash).}
			\label{fig7}
		\end{figure}

		We study in \cref{fig7} the relationship between the fitness and the environmental fluctuation timescale, $\theta$, for a fixed phenotype switching strategy.  In \cref{fig7}a,b we see through the numerical solution of the Fokker-Planck equation convergence to the perfect correlation between phenotype composition and the environment, denoted $x = \tilde{x}(z)$. In \cref{fig7}c,d we compare the fitness calculated from the Fokker-Planck equation (which is valid only for $\theta$ sufficiently large) to repeated simulation of the coupled SDE system, and an approximation based on a moment closure solution to the SDE system \cite{Sukys.2021}. We observe in all three methodologies a monotonic relationship between fitness and $\theta$ that connects the slow regime to the fast regime as $\theta$ increases.  Further, it follows directly from \cref{fitness_gentheta} that the fitness as $\theta \rightarrow \infty$, for all environments, is lower than that for a homogeneous strategy. 
			
	While we are not able to prove that the relationship between fitness and $\theta$ is monotonic, we can see that a sufficient (albeit not necessary) condition for existence of a regime where heterogeneity is advantageous is simply that the fitness under the QSS approximation is greater than that for the homogeneous strategy. Two further advantages of studying the QSS regime are numerical tractability, as computation of the fitness for the two-phenotype case (\cref{fitness_qss}) involves only the numerical calculation of an integral, and tractability for a more general (potentially continuously distributed) population structure. 

\begin{figure}[!t]
		\centering
		\includegraphics{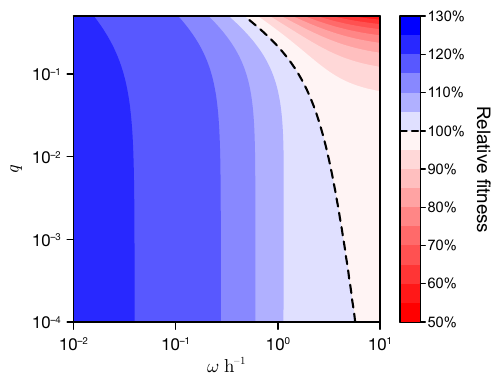}
		\caption[Fig 8]{\textbf{Quasi-steady-state fitness for a discrete strategy in a continuously fluctuating Gamma environment.} We calculate the slow environment limit of the fitness as a function of the target dormant proportion, $q$, and the switching rate, $\omega$. Fitness shown is relative to that for the homogeneous strategy. }
		\label{fig8}
	\end{figure}

	QSS results in \cref{fig8} show the fitness of a heterogeneous strategy with a general target distribution, $q$, and switching rate, $\omega$, relative to a homogeneous strategy. The most obvious result is that heterogeneity is not evolutionarily stable for large switching rates: an analogous result was observed by \cite{Thattai:2004}, who study responsive adaptation of bacteria to switching (Poisson-like) environments. Secondly, we see that fitness is a decreasing function of the switching rate, paradoxically suggesting that a heterogeneous population that does not permit phenotype switching (and therefore, cannot regulate heterogeneity) has the highest fitness. However, implicit in the QSS is the assumption that cellular dynamics rate parameters are sufficiently separated from that of the environment, such that $\omega \gg \theta$. Therefore, we take our results in \cref{fig8} to demonstrate the existence of regimes in which a heterogeneous strategy is advantageous, rather than a demonstration of the optimal or evolutionarily stable switching strategy, which is only defined for finite $\theta$.
	
	\subsubsection{Bivariate continuously distributed phenotypes}
	
	We next consider that a  two-phenotype population  develops continuous heterogeneity such that the target distribution is given by the Gaussian mixture in \cref{discrete_mixture}. We again consider a perturbation expansion in the QSS fitness,
		\begin{equation}\label{biv_cont_expansion}
			\varphi_\mathrm{QSS} = \mathbb{E}_z\mathbb{E}_{\tilde{p}(\varepsilon,z)}\big(\lambda(\varepsilon,z)\big) \sim c_0 + c_1 \eta_1 + c_2 \eta_2,
		\end{equation}
	such that $c_1 > 0$ indicates a fitness advantage to the population in employing continuous heterogeneity in the neighbourhood of $\varepsilon = 0$, and $c_2 > 0$ an analogous fitness advantage near $\varepsilon = 1$. 
	
	Full details of the methodology are provided in the supplementary material. In summary, we construct the perturbation expansion by considering that $\tilde{p}(\varepsilon,z)$ can be, for fixed $z$, represented as a mixture
		\begin{equation}
			\tilde{p}(\varepsilon,z) = \beta \tilde{p}_0(\varepsilon,z) + (1 - \beta) \tilde{p}_1(\varepsilon,z)
		\end{equation}
	where the variance of each component is of order $\eta_1$ and $\eta_2$, respectively.  Substituting into the governing equation for $\tilde{p}$, \cref{qss1},  taking a moment expansion \cite{Browning.2022}, and assuming that higher-order central moments vanish, we obtain a leading-order approximation for the moments of $\tilde{p}_i(\varepsilon,z)$ in terms of $\eta_1$ and $\eta_2$. The coefficients in \cref{biv_cont_expansion} can then be calculated by numerically integrating over $z$. 
	
	\begin{figure}[!t]
		\centering
		\includegraphics{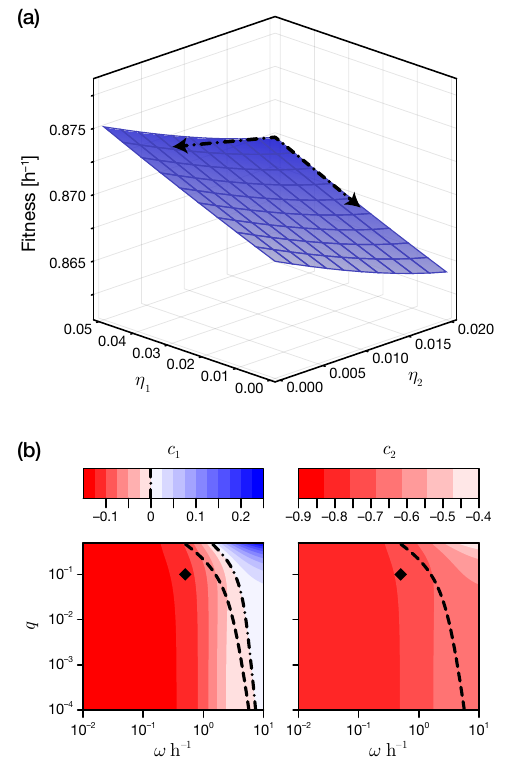}
		\caption[Fig 9]{\textbf{Bivariate continuous heterogeneity in a continuously fluctuating Gamma environment.} (a) We calculate the fitness using a QSS approximation to the equilibrium phenotype composition (blue surface) and a perturbation expansion (black dashed). (b) The perturbation coefficients $c_1$ and $c_2$ (of $\eta_1$ and $\eta_2$, respectively) as a function of $q$ and $\omega$. The dash-dot line indicates the boundary between regions with negative (red) and positive (blue) $c_1$. Dashed lines correspond to that in \cref{fig8}; regions to the left are those in which the discrete heterogeneous strategy is advantageous.}
		\label{fig9}
	\end{figure}
	
	Results in \cref{fig9}a show the fitness for the Gamma environment studied in \cref{fig7}d, for general $\eta_1$ and $\eta_2$, in addition to the perturbation expansion for $\eta_i \ll 1$. The availability of a numerically tractable perturbation expansion allows us to establish regimes where continuous heterogeneity is advantageous by investigating the coefficients $c_1$ and $c_2$ as functions of the strategy $(q,\omega)$. In \cref{fig9}b we show that $c_2$ is, for the environment considered, always negative (i.e., it is never evolutionarily stable to present with continuous heterogeneity around the phenotype with the highest average fitness), while $c_1$ is negative for all but sufficiently large $\omega$ and $q$.  However, $c_1 > 0$ is only the case  in regions where the heterogeneous strategy is itself disadvantageous; thus, continuous heterogeneity near $\varepsilon = 0$ gives a fitness advantage only in the case that the existence of a phenotype at $\varepsilon = 0$ is itself disadvantageous.

	
	\subsection{Fluctuations in the fittest phenotype}
	
	We have, thus far, restricted our study to growth rates that are monotonic in the phenotype index. Such environments are typically considered in the context of bet-hedging: the single optimal phenotype is often on the edge of the phenotype space, and the evolutionary stable strategy involves regulating a proportion of cells at the two extrema. We now extend this analysis to a scenario in which the environment induces changes in the optimal phenotype itself to represent, for example, temperature specialisation of bacteria in  environments in which the temperature fluctuates continuously \cite{Charlebois.2018,Nguyen.2021,Lambros.2021}. 
	
	\begin{figure*}[!t]
		\centering
		\includegraphics{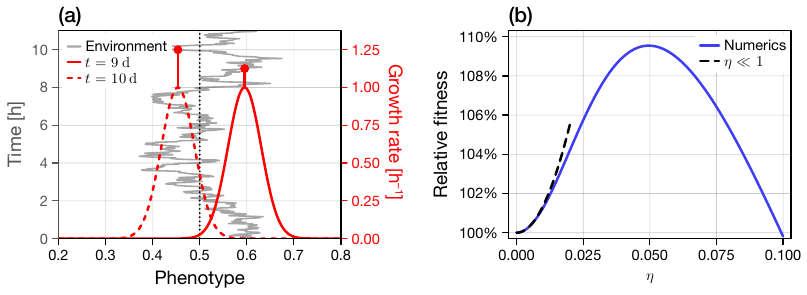}
		\caption[Fig 10]{\textbf{Continuously fluctuating optimal phenotype.} (a) The growth rate of each phenotype is given by a Gaussian centered at the current value of the environment variable $z(t)$, modelled as an Ornstein-Uhlenbeck process with mean $\varepsilon = 0.5$. Time shown on the left axis, with growth rate at $t = \SI{9}{\hour}$ and $t = \SI{10}{\hour}$ is shown on the right axis. The phenotype $\varepsilon = 0.5$ is the optimal phenotype if a homogeneous strategy is employed. (b) Fitness (relative to a homogeneous strategy) calculated for a continuous target distribution with variance $\eta^2$, calculated numerically (blue) and using a perturbation expansion (black dashed). Other parameters are given by $\xi = 0.5$, $k = \SI{1}{\per\hour}1$, $\sigma = 0.1$.}
		\label{fig10}
	\end{figure*}
		
	We again assume that environmental fluctuations are driven by an Ornstein-Uhlenbeck process (\cref{zou}), relating the environment to the growth rate more directly through
		\begin{equation}
			\lambda(\varepsilon,w) = k \exp\left(\dfrac{-(\varepsilon - w)^2}{\xi^2}\right),
		\end{equation}
	where $w = \tfrac{1}{2} + \sigma z \sim \mathcal{N}(0.5,\sigma)$ describes the (fluctuating) location of the optimal phenotype. In \cref{fig10}a, we show a realisation of the environment, demonstrating the time-dependent coupling between the environment and the growth rate of every phenotype. While many other choices of $\lambda(\varepsilon,z)$ are reasonable, our specific choice enables us to make analytical progress toward a perturbation solution for the fitness in the case that the population develops continuous heterogeneity around an otherwise optimal singular phenotype. We also note that, as the governing equation for the phenotype composition depends only on a difference in growth rates, we can, without loss of generality, restrict our study to functions such that $\min \lambda(\varepsilon,w) = 0$. 

	Due to symmetry, the single optimal phenotype (i.e., in a homogeneous strategy) is located at $\varepsilon^* = \tfrac{1}{2}$. We again consider that the population undergoes mutations that yield a continuously distributed target distribution $\hat{p}(\varepsilon) = \mathcal{N}(\varepsilon^*,\eta)$. Substituting into \cref{qss1}, assuming that $\eta \ll 1$, and integrating over $z$, we find that the fitness is given by	
		\begin{equation}\label{fitness_fluct_optima}
			\varphi \sim \dfrac{k \xi}{\sqrt{\xi^2 + 2\sigma^2}} + \left(\dfrac{4k^2 \sigma^2}{\xi \omega (\xi^2 + 4\sigma^2)^{3/2}} - \dfrac{k\xi}{(\xi^2 + 2\sigma^2)^{3/2}} \right) \eta^2 + \mathcal{O}(\eta^3).
		\end{equation}
	For a specific choice of parameters, we demonstrate in \cref{fig10}b that the expansion conforms with a numerical solution generated for general $\eta$. Both the expansion and the numerical results clearly demonstrate the existence of a regime in which continuous heterogeneity is advantageous. While the relationship between fitness and $\eta$ is clearly not always monotonic, the positivity of the correction term in \cref{fitness_fluct_optima} gives a sufficient condition for the existence of an evolutionarily stable heterogeneous strategy. Furthermore, we can see that heterogeneity is immediately (i.e., for small $\eta$) disadvantageous in cases where either $\omega \gg 0$, if $\sigma \rightarrow 0$, or if $\xi \rightarrow 0$.

\FloatBarrier
\section{Discussion  and Conclusions }

	Many biological systems regulate and exploit phenotypic heterogeneity to achieve functional aims \cite{Elsasser.1984uy,Kussell.2005v2}. The phenomenon of bet-hedging, in particular, is highly studied, particularly through the lens of a discrete set of highly specialised phenotypes \cite{Balaban:2004,Kussell.2005,Jolly.2018}. Ecological theory commonly describes diversification strategies as an evolutionary response to environmental uncertainty \cite{Muller.201303s}. In this context, experimental and mathematical models of environmental fluctuations are predominantly characterised by an effective growth rate that switches either periodically or stochastically between a finite set of conditions: a growth-promoting environment, and a stressful environment, for example \cite{Kussell.2005,Acar.2008}. More general and arguably more realistic models of environment are comparatively less studied.  Our goal in the present work is to develop a mathematical framework to demonstrate the existence of regimes where phenotypic heterogeneity is advantageous as an evolutionary response to a set of generalisable stochastic environments. 

	\subsection{Premise and modelling assumptions}
	
	We restrict our analysis to a simple cellular dynamics model comprising heritable traits and an exponentially growing, well-mixed, population; the latter motivated by serial-dilution or continuous-culture experiments where the population density is kept sufficiently low that both cell-to-cell competition and nutrient consumption are negligible. We do not explicitly model evolution. Rather, our premise is that, in a given stochastically fluctuating environment, populations have evolved a fixed phenotype switching strategy that regulates a heterogeneous composition of phenotypes. We consider a Malthusian-like measure of fitness corresponding to the expected per-capita growth rate. Strategies with a locally maximal fitness are, therefore, considered to be potentially evolutionarily stable. Determination of the fitness is, in general, analytically intractable. We are, therefore, primarily concerned with demonstrating environments and heterogeneous switching strategies carry a fitness advantage over a single-phenotype homogeneous strategy.

     \subsection{Environments in which discrete heterogeneity is advantageous}
    
	 We first establish regimes in which the standard two-phenotype model of bet-hedging is advantageous. For temporally uncorrelated environments, we derive a simple set of analytical expressions (\crefrange{whitenoise_cond1}{whitenoise_cond2}) that indicate environmental regimes in which a population can increase its fitness by employing a second phenotype with lower mean growth rate. These regimes become more restrictive as the correlation between the inter-phenotype growth rates tends to unity, and in the case that the growth rate of the fittest phenotype has zero variance (i.e., when growth is deterministic). Conversely, regimes become less restrictive if growth rates are un- or inversely-correlated; representative, for example, of regimes in which phenotypes specialise to different nutrient sources that fluctuate independently or inversely (the latter arising where a fixed \textit{total} amount of nutrient is available). 
			
	
	 Environments that are temporally correlated are, in general, neither analytically or numerically tractable. We make progress by studying fitness in what we term the \textit{slow regime}: that is, the scenario where environmental changes occur on a much slower timescale to that of the cellular dynamics. This allows us apply a QSS assumption, whereby the phenotype composition is assumed to be at equilibrium and, therefore, deterministically coupled to the environment. This assumption is likely most appropriate in environments where changes are continuous (i.e., environments driven by Ornstein-Uhlenbeck processes studied in \cref{fig7}): the phenotype composition will always remain comparatively close to its equilibrium. The well-studied Poisson environment, on the other hand, is less conducive to the QSS assumption as growth rates transition instantaneously. For all environments, we are able to show that heterogeneity is never advantageous in environments that fluctuate on very fast timescales: the \textit{fast regime}. A sufficient but not necessary condition for heterogeneity being advantageous temporally correlated environments is, therefore, that it is advantageous in the slow regime.
	
	Our results in \cref{fig7} demonstrate a set of environments in which a two-phenotype strategy is advantageous, and show a monotonic relationship between the timescale of environmental switching and the fitness of a two-phenotype strategy. Applying the QSS assumption in \cref{fig8} reveals a balance between the timescales of phenotype switching and environmental changes. In particular, the heterogeneous strategy becomes disadvantageous in populations that present fast switching between phenotypes. Advantages are gained, therefore, in regimes where phenotype switching is sufficiently slow, but not necessarily as slow as that of the environmental timescales in which the QSS assumption would apply. These somewhat paradoxical results are also seen in \cite{Thattai:2004}, and can be explained by considering the antithetical scenario that environmental fluctuations occur very quickly (the fast regime). In such a regime, the environment and population structure will be statistically independent (which is also the case if switching is sufficiently fast), and it is straightforward to see from \cref{fitness_qss} that the fitness of a two-phenotype population will be lower than that with an optimised single phenotype.

    \subsection{Environments in which bimodal continuous heterogeneity is advantageous}
 
	 While study of bet-hedging strategies that comprise a discrete set of phenotypes is extensive, there is comparatively very little theory to allow the study of a continuous distribution of phenotypes. While factors such as gene expression noise mean that within-subpopulation variability is likely omnipresent, it is routine for theoretical models of bet-hedging to assume that cells within each subpopulation (e.g., dormant or proliferative) behave identically.  Our most important result is a demonstration of regimes in which subpopulation heterogeneity is evolutionarily stable. In particular, we apply a perturbative approach to study the effect of small within-subpopulation variance on fitness in two main scenarios: first, that an otherwise optimal homogeneous population develops heterogeneity (unimodal), and secondly that continuous heterogeneity develops in an initially two-phenotype population (bimodal). For environments driven by white-noise processes, we again establish clear parameter regimes in which continuous heterogeneity is advantageous (\crefrange{whitenoise_cont_1}{whitenoise_cont_4}). As our asymptotic results are dependent only on the local behaviour of the growth rate variance and correlations, we expect our findings to hold for all differentiable choices of these functions. Results for the bimodal scenario are less analytically interpretable, however we demonstrate in \cref{fig4} that, for a particular choice of parameters, only continuous heterogeneity in the dormant phenotype is advantageous. 

	In temporally correlated environments, analogous perturbative correction terms even in the slow regime are analytically intractable. We do, however, derive semi-analytical expressions that allow numerical computation of these corrections terms, and therefore allow us to identify environments in which continuous heterogeneity may be advantageous. For a particular choice of environment, results in \cref{fig9} show that subpopulation variance is always immediately disadvantageous. This does not rule out a heterogeneous strategy that is advantageous for a significantly high variance, as seen in \cref{fig3}c for an environment driven by white-noise. These results have further parallels to our findings for temporally uncorrelated environments: semi-analytical expressions for the correction terms (supplementary material) suggest that a necessary condition for a positive correction at the proliferative state may be a negative correction at the dormant state. An aspect of continuous heterogeneity that we have not considered is one that arises in some sense extrinsically out of gene expression noise. That is, a discretely structured population may be evolutionarily favoured \textit{in theory}, however can only be regulated at the cost of some level of subpopulation variance. Further study using our perturbative approach is likely to elucidate the cost-benefit trade-off for bet-hedging in discretely structured populations.

    \subsection{Unimodal continuous heterogeneity}

	The classical view of bet-hedging is risk-spreading: much of the present work has involved environments that present risk to a homogeneous strategy through periods of time with negative growth rates. Our final result (\cref{fig10}), demonstrates that continuous unimodal heterogeneity can itself be advantageous in the absence of an explicit stressor if the environment drives fluctuations in the optimal phenotype. Again applying the perturbative approach, we provide for a specific growth rate model an analytically tractable expression for the correction term (\cref{fitness_fluct_optima}), providing a sufficient condition for an evolutionary stable continuous subpopulation structure. One motivation for this scenario may be that of temperature specialisation in populations of bacteria subject to relatively slow fluctuations in temperature \cite{Charlebois.2018}. The optimal homogeneous strategy is to specialise to the mean temperature, while our results show that the population can increase its fitness by introducing small amounts of phenotypic variability in which isogenic cells specialise to different temperatures.

    \subsection{Future work}

	A key theoretical and experimental challenge that we have not considered is that bet-hedging strategies are likely to vary significantly between growth and stationary phase. In multi-resource environments where nutrients are limited, for example, populations cannot grow exponentially and frequency dependent selection may lead to diversification strategies that include otherwise less fit phenotypes \cite{Healey.2016}. A body of relatively recent work explores fluctuations in environmental carrying capacity (rather than growth rate), and could, in principle, be extended to allow for a continuous distribution of phenotypes in more general environments \cite{Wienland.2018,Taitelbaum.2020,Taitelbaum.2023,Hernandez-Navarro.2023}. Another aspect that we have not considered is the role of spatial fluctuations, particularly in the context of nutrient depletion. Extending our analysis to explicitly incorporate fluctuations in both space and time is likely to render the model intractable. However, our results may still provide insight in the context of low density motile populations in spatially heterogeneous environments: from the perspective of individual cells, the nutrient concentration will, effectively, fluctuate temporally.   
	
	The formulation of a more biologically realistic model of phenotype regulation is likely to be driven by future experimentation. In an exponentially growing population within continuous culture, it is clear that active phenotypic switching is required to avoid extinction of subpopulations that are overall less fit. In the cancer modelling literature, it is typical for transitions between phenotypes to be local and modelled as diffusion through a (potentially constrained) phenotype space \cite{Chisholm.2015}.   At an individual level, local changes in phenotype can also be captured using an Ornstein-Uhlenbeck process that reverts to a potentially time-varying optimal phenotype \cite{DeSouzaSilva.2023,Browning.2024dzf}. Our modelling approach provides a framework in which to experimentally test subpopulation-regulation models, by predicting advantageous phenotype compositions for arbitrarily (albeit slowly) fluctuating environments. In most cases, however, a phenotype is likely to result from the expression of a large number of genes, and may manifest as a high-dimensional distribution of behaviours \cite{Doebeli.2017}. In the slow regime through the QSS approximation, at least, we anticipate that our theory can be extended to both two-dimensional phenotype spaces and to environmental fluctuations driven by two-dimensional stochastic processes.

    \subsection{Concluding remarks}
    
	It is fast becoming understood that heterogeneity is not only ubiquitous to biology, but plays a key role in the function of many biological systems. We elucidate this function in populations exposed to a range of stochastically fluctuating environments. Importantly, we demonstrate regimes in which regulation of both discrete and continuous heterogeneity provides a fitness advantage to populations. Overall, our work enables future theoretical and experimental work to investigate and eventually exploit phenotypic heterogeneity in biologically realistic environments.

\section*{Data availability}

Code used to produce the results is available at \url{https://github.com/ap-browning/heterogeneity}.

\section*{Acknowledgements}

APB thanks the Mathematical Institute, University of Oxford, for a Hooke Research Fellowship. SH was funded by Wenner-Gren Stiftelserna/the Wenner-Gren Foundations (WGF2022-0044) and the Kjell och M{\"a}rta Beijer Foundation. 


\end{document}


	\title{Supplementary material for\\``Heterogeneity in temporally fluctuating environments''}


	\author[1,2*]{Alexander P Browning}
	\author[3]{Sara Hamis}
	
	\affil[1]{School of Mathematics and Statistics, University of Melbourne, Melbourne, Australia}
	\affil[2]{Mathematical Institute, University of Oxford, Oxford, United Kingdom}
	\affil[3]{Department of Information Technology, Uppsala University, Uppsala, Sweden}
	
	\date{\today}


	\maketitle
	\tableofcontents
	\footnotetext[1]{Corresponding author: alex.browning@unimelb.edu.au}

\clearpage
\section{White-noise environments}

	\subsection{Binary model}

	The growth rates and target distribution are given by
		\begin{align*}
			\lambda(\varepsilon) &= \left\{\begin{array}{ll}
				\mu_1 + \sigma_1 \xi_1, & \varepsilon = 1,\\
				\mu_2 + \sigma_2 \rho \xi_1 + \sigma_2 \sqrt{1 - \rho^2} \xi_2, & \varepsilon = 0,
				\end{array}\right.\\
			p(\varepsilon) &= \left\{\begin{array}{ll}
				1 - q 	& \varepsilon = 1,\\
				q 		& \varepsilon = 0.
				\end{array}\right.
		\end{align*}

	Thus, the fitness is given by
		\begin{align*}
			\varphi &= \mathbb{E}_{p(\varepsilon)}\big(\mu(\varepsilon)\big) - \dfrac{1}{2}\sum_{i=1}^2 \mathbb{E}_{p(\varepsilon)}^2 \big(s_i(\varepsilon)\big),\\
				&= \mu_1 - \dfrac{\sigma_1^2}{2} + q (\mu_2 - \mu_1 + \sigma_1^2 - \rho \sigma_1 \sigma_2) + q^2\left(\rho \sigma_1\sigma_2 - \dfrac{\sigma_1^2 + \sigma_2^2}{2} \right),
		\end{align*}
	which has a critical point at
		\begin{equation*}
			q^* = \dfrac{\mu_2 - \mu_1 + \sigma_1^2 - \rho \sigma_1 \sigma_2}{\sigma_1^2 + \sigma_2^2 - 2\rho \sigma_1\sigma_2}.
		\end{equation*}
	
	\subsection{Binary model (finite switching speed)}

	We first apply It\^o's lemma to obtain an SDE for the proportion of the population in the dormant phenotype, $p(0,t) := x(t)$, given by
		\begin{subequations}
		\begin{align*}
			\dd x &= f_x(x) \, \dd t + g_x(x) \, \dd W,\\
			f_x(x) &= b - a x - bx - x(1-x)(\mu_1 - \mu_2 - (1 - x)\sigma_1^2 + (1 - 2x)\rho \sigma_1\sigma_2 + x\sigma_2^2,\\
			g_x(x) &= x(1-x)\sqrt{\sigma_1^2 - 2\rho\sigma_1\sigma_2 + \sigma_2^2},
		\end{align*}
		\end{subequations}
	for $a = \omega (1 - q)$ and $b = \omega q$.
		
	The Fokker-Planck equation governing the equilibrium distribution of $x$ is given by
		\begin{equation*}
			0 = -\pdv{\big(f_x(x) u(x)\big)}{x} + \dfrac{1}{2}\pdv[2]{\big(g_x^2(x) u(x)\big)}{x},
		\end{equation*}
	subject to the usual boundary conditions that $u(0) = u(1) = 0$, and that $\int_0^1 u(x) \,\dd x = 1$. 
	
	Integrating once we obtain
		\begin{equation*}
			0 = -f_x(x) u(x) + \dfrac{1}{2}\pdv{\big(g_x^2(x) u(x)\big)}{x},
		\end{equation*}
	which we solve using integrating factor to arrive at
		\begin{equation}\label{fp1sol}
			u(x) \propto \dfrac{1}{g^2(x)} \exp\left(2\int_0^x \dfrac{f_x(s)}{g_x^2(s)} \,\dd s \right).
		\end{equation}
	The integral in the exponent of \cref{fp1sol} can be computed analytically, yielding 
		\begin{equation*}
			u(x) = \dfrac{C}{g^2(x)} \mathrm{e}^{\alpha(x)},
		\end{equation*}
	where
		\begin{equation*}
			\alpha(x) = \dfrac{2x \left(\mu_1-\mu_2-\sigma_1^2+\rho \sigma_1 \sigma_2\right)+x^2\left(\sigma_1^2-2 \rho \sigma_1 \sigma_2+\sigma_2^2\right)- 2a \log(1-x)- 2b \log(x)}{\sqrt{\sigma_1^2 - 2\rho\sigma_1\sigma_2 + \sigma_2^2}},
		\end{equation*}
	and where $C$ is a normalisation constant to be determined using quadrature to ensure that $u(x)$ is a valid probability density function.

	\subsection{Unimodal continuous heterogeneity}\label{whitenoise_unimodal}

	We here assume that $p(\varepsilon) = \mathcal{N}_{(0,1)}(\varepsilon^*,\eta^2)$ for $\eta \ll 1$ and wish to calculate the fitness
		\begin{equation}\label{whitenoise-fitness}
			\varphi = \mathbb{E}_{p(\varepsilon)}\big(\mu(\varepsilon)\big) - \dfrac{1}{2}\sum_{i=1}^2 \mathbb{E}_{p(\varepsilon)}^2 \big(s_i(\varepsilon)\big),
		\end{equation}
	in the case that $\mu(\varepsilon) = \varepsilon$ and where
		\begin{subequations}\label{s12}
		\begin{align}
			s_1(\varepsilon) &= \rho(\varepsilon - \varepsilon^*)\sigma(\varepsilon),\\
			s_2(\varepsilon) &= \sigma(\varepsilon)\sqrt{1 - \rho^2(\varepsilon - \varepsilon^*)}.
		\end{align}	
		\end{subequations}

	For $\eta$ sufficiently small, we can consider firstly that $s_1$ and $s_2$ are well approximated in the vicinity of $\varepsilon^*$ by their respective Taylor expansions, and secondly that $p(\varepsilon)$ is very well approximated by an untruncated distribution (in the case that $\varepsilon^* = 1$, an untruncated half Gaussian distribution). Together, these assumptions yield analytical expressions for the respective expectations in \cref{whitenoise-fitness} (see Mathematica code).
	
	Care must be taken for the case that $\varepsilon^*$ is on the interior of the domain and in which $\rho(\cdot)$ is not differentiable: in this case, we note that $\rho(\cdot)$ is symmetric, and thus consider a piecewise formulation of \cref{s12}.
		\begin{subequations}
		\begin{align*}
			s_1(\varepsilon) &= \left\{\begin{array}{ll}
				\rho_-(\varepsilon - \varepsilon^*)\sigma(\varepsilon) & \varepsilon \le \varepsilon^*,\\
				\rho_-(\varepsilon^* - \varepsilon)\sigma(\varepsilon) & \varepsilon > \varepsilon^*,\\
			\end{array}\right.,\\
			s_2(\varepsilon) &= \left\{\begin{array}{ll}
				\sigma(\varepsilon)\sqrt{1 - \rho_-^2(\varepsilon - \varepsilon^*)} & \varepsilon \le \varepsilon^*,\\
				\sigma(\varepsilon)\sqrt{1 - \rho_-^2(\varepsilon^* - \varepsilon)} & \varepsilon > \varepsilon^*,\\
			\end{array}\right.,
		\end{align*}	
		\end{subequations}
	where $\rho_-(\Delta)$ gives the correlation function for $\varepsilon \le \varepsilon^*$. 

	\clearpage
	\subsection{Bimodal continuous heterogeneity}

	We proceed in a similar manner to in \Cref{whitenoise_unimodal}, however for a slightly more general model without the same restrictions on $\mu(\varepsilon)$ (to enable a direct comparison to the results in Fig. 2 of the main document), although here we only present the expansion for a form of $\rho(\cdot)$ that is differentiable at $\Delta = 0$. 
	
	The population composition is given by the mixture
		\begin{equation*}
			p(\varepsilon) = q \mathcal{N}_{(0,1)} (0,\eta_1) + (1 - q) \mathcal{N}_{(0,1)} (1,\eta_2),
		\end{equation*}
	such that the population tends to a discrete binary structure for $\eta_1, \eta_2 \rightarrow 0$. The required expectations in the fitness (given by \cref{whitenoise-fitness}) are calculated for each component of the mixture seperately under the assumption that, for $\eta_1$ sufficiently small, the density of $\mathcal{N}_{(0,1)}(0,\eta_1)$ is approximately that of a half Gaussian distribution (and similar for the distribution near $\varepsilon = 1$ for $\eta_2 \ll 1)$. Full details are given in the Mathematica code.

\section{Poisson environment}

	\subsection{Binary model (Fokker-Planck)}

	We again denote the current proportion of cells in the dormant state ($\varepsilon = 0$) by $x(t)$, with dynamics given by
	\begin{equation}\label{poisson_fx}
		\dv{x}{t} = f_x(x,\lambda(z)) = b - x (a - b - \lambda(z)(1-x))	
	\end{equation}
	with $a = \omega (1 - q)$, $b = \omega q$, and where $\lambda(z)$ gives the growth rate of cells at $\varepsilon = 1$ (cells at $\varepsilon = 0$ are assumed to be completely dormant). The environment is given by a homogeneous Poisson process with rates given by $s_1 = \theta / (1 - \alpha)$ and $s_2 = \theta / \alpha$ for the growing and stressful states respectively (in which the growth rates are $\lambda(1) = \lambda_1$ and $\lambda(-1) = \lambda_2$, respectively). 
	
	The joint stationary probability density $u(x,\lambda)$ is governed by the coupled system of partial differential equations
		\begin{equation}\label{poiss_pde1}
		\begin{aligned}
			\pdv{}{x}\Big(f_x(x) u(x,\lambda_1)\Big) &= -s_1 u(x,\lambda_1) + s_2 u(x,\lambda_2),\\
			\pdv{}{x}\Big(f_x(x) u(x,\lambda_2)\Big) &= \phantom{-}s_1 u(x,\lambda_1) - s_2 u(x,\lambda_2),
		\end{aligned}
		\end{equation}
	subject to the constraint that $\int_0^1 u(x,\lambda_1) + u(\lambda_2) \,\dd x = 1$. 
	
	The lack of a diffusion term in \cref{poiss_pde1} renders the problem numerically unstable. We thus regularise the problem by introducing a small diffusion term such that the dynamics of $u(x,\lambda(z))$ are given, approximately, by
		\begin{equation}\label{poiss_pde2}
		\begin{aligned}
			\pdv{}{x}\Big(f_x(x) u(x,\lambda_1)\Big) &= -s_1 u(x,\lambda_1) + s_2 u(x,\lambda_2) + \beta\pdv[2]{u(x,\lambda_1)}{x},\\
			\pdv{}{x}\Big(f_x(x) u(x,\lambda_2)\Big) &= \phantom{-}s_1 u(x,\lambda_1) - s_2 u(x,\lambda_2) + \beta\pdv[2]{u(x,\lambda_2)}{x},
		\end{aligned}
		\end{equation}
	where $\beta$ is small (specifically, we set $\beta = 10^{-4}$). We then solve \cref{poiss_pde2} numerically using a finite volume discretisation alongside a linear solver\footnote{\href{https://github.com/ap-browning/heterogeneity/blob/main/library/poisson/solve_binary_poisson.jl}{\texttt{library/poisson/solve\_binary\_poisson.jl}}}. The fitness is then given by
		\begin{equation*}
			\varphi = \sum_{i=1}^2\int_0^1 \lambda_i (1 - x) u(x,\lambda_i) \,\dd x,
		\end{equation*}
	which we calculate using quadrature\textsuperscript{1}. 

	\subsection{Binary model (QSS and fast regimes)}\label{qssbinary}

	At QSS, we assume in \cref{poisson_fx} that $\dd x/\dd t = 0$. This yields an expression for $x$ in terms of $\lambda(z)$, denoted $\tilde{x}(\lambda)$:
		\begin{equation*}
			\tilde{x}(\lambda) = \dfrac{-a + b + \lambda + \sqrt{4b\lambda + (-a + b + \lambda)^2}}{2\lambda}.
		\end{equation*}
	The fitness is then given by
		\begin{equation*}
			\varphi = \sum_{i=1}^2 (1 - \tilde{x}(\lambda_i)) \lambda_i u(\lambda_i),
		\end{equation*}
	where $u(\lambda_i)$, the probability mass function for the environment, is given simply by
		\begin{equation*}
			u(\lambda) = \left\{\begin{array}{ll}
				1 - \alpha & \lambda = \lambda_1,\\
				\alpha & \lambda = \lambda_2.
			\end{array}\right.
		\end{equation*}

	In the fast regime, $x$ and $\lambda$ are uncorrelated; thus, we take $\lambda$ to be at its mean value, given by
		\begin{equation*}
			\bar\lambda = \sum_{i=1}^2 \lambda_i u(\lambda_i),
		\end{equation*}
	such that $x$ is given by $\tilde{x}(\bar\lambda)$. The fitness is given simply by
		\begin{equation*}
			\varphi = (1 - \tilde{x}(\bar\lambda))\bar\lambda.
		\end{equation*}
	%

\section{Continuously fluctuating environments}

	\subsection{Binary model (Fokker-Planck)}

	The coupled dynamics of $x(t)$ and $z(t)$ are given by the system of SDEs
		\begin{equation}
		\begin{aligned}
			\dd x(t) &= f_x(x,\lambda(z))\; \dd t + \sqrt{2\beta} \dd W_1,\\
			\dd z(t) &= -\theta z(t) \; \dd t + \sqrt{2\theta} \dd W_2,
		\end{aligned}
		\end{equation}
	where we have added an artificial diffusion term of magnitude $\beta$ to regularise the problem, and where
		\begin{equation*}
			f_x(x,\lambda(z)) = b - x \Big(a - b - \big(\lambda(1,z) - \lambda(0,z)\big)(1-x)\Big).
		\end{equation*}

	The joint stationary distribution, $u(x,z)$, is given by the solution to the two-dimensional Fokker-Planck equation
		\begin{equation}\label{ou_fp1}
			\pdv{}{x}\big(f_x(x,z) u(x,z)\big) - \theta \pdv{}{z}\big(z u(x,z)\big) = \beta \pdv[2]{u(x,z)}{x} + \theta \pdv[2]{u(x,z)}{z},
		\end{equation}
	subject to the condition that $u(x,z)$ must be a valid probability density function.
	
	We solve \cref{ou_fp1} numerically using a linear finite volume discretisation, combined with an iterative solve for the resultant fast linear system (GMRES)\footnote{\href{https://github.com/ap-browning/heterogeneity/blob/main/library/ornstein-uhlenbeck/solve_binary_ou.jl}{\texttt{library/ornstein-uhlenbeck/solve\_binary\_ou.jl}}}. The fitness is then given by
		\begin{equation}
			\varphi = \int_{-\infty}^\infty \int_0^1 \bigg(x \lambda(0,z) + (1 - x) \lambda(1,z)) u(x,z) \,\dd x \dd z,
		\end{equation}
	which we calculate numerically using quadrature\textsuperscript{2}.

	\subsection{Bimodal continuous fluctuations}

	The target distribution is given now by the mixture
		\begin{equation*}
			\hat{p}(\varepsilon) = q \hat{p}_1(\varepsilon) + (1 - q) \hat{p}_2(\varepsilon),
		\end{equation*}
	where $\hat{p}_1(\varepsilon)$ is the density function for the truncated Gaussian $\mathcal{N}_{(0,1)}(0,\eta_1)$, and $\hat{p}_2(\varepsilon)$ that for $\mathcal{N}_{(0,1)}(1,\eta_2)$.
	
	At QSS, the population composition distribution is given by the solution to
		\begin{equation}\label{qss1}
			0 = \tilde{p}(\varepsilon,z)\Big[\lambda(\varepsilon,z) - \mathbb{E}_{\tilde{p}(\varepsilon,z)} \big(\lambda(\varepsilon,z)\big) - \omega\Big] + \omega \hat{p}(\varepsilon),
		\end{equation}
	in which we are primarily interested in (for a fixed $z$, or equivalently, a fixed $\lambda$) the quantity $\mathbb{E}_{\tilde{p}(\varepsilon,z)} \big(\lambda(\varepsilon,z)\big)$. Given the structure of the target distribution as a mixture, our approach is to assume that $\tilde{p}(\varepsilon,z)$ is also given by the mixture
		\begin{equation}\label{ptildemix}
			\tilde{p} = \gamma \tilde{p}_1 + (1 - \gamma)\tilde{p}_2,
		\end{equation}
	where $\gamma$ is also to be determined, and where, for brevity, we drop the notation giving function arguments of density functions explicitly. By definition, we assume that the mass of $\tilde{p}_1$ is concentrated around $\varepsilon = 0$ (specifically, within the support of $p_1$), and similar for $\tilde{p}_2$ near $\varepsilon = 1$. 
	
	Before proceeding, we define the following quantities
		\begin{subequations}
		\begin{align*}
			&&E_1 &= \mathbb{E}_{\tilde{p}_1}(\varepsilon) \sim \mathcal{O}(\eta_1), & E_2 &= \mathbb{E}_{\tilde{p}_2}(1 - \varepsilon) \sim \mathcal{O}(\eta_2),&&\\
			&&V_1 &= \mathbb{E}_{\tilde{p}_1}(\varepsilon^2) \sim \mathcal{O}(\eta_1^2), & V_2 &= \mathbb{E}_{\tilde{p}_2}((1 - \varepsilon)^2) \sim \mathcal{O}(\eta_2^2),&&
		\end{align*}	
		\end{subequations}
	where the order follows from the assumption that $\tilde{p}_i$ will remain within the support of $\hat{p}$. 

	We will soon determine the leading order asymptotic behaviour of $E_1$ and $E_2$, from which we can calculate the expected growth rate by considering first that (for a fixed $z$)
		\begin{equation}\label{Eplambda}
			\mathbb{E}_{\tilde{p}} \big(\lambda(\varepsilon,z)\big) = \gamma \mathbb{E}_{\tilde{p}_1}(\lambda(\varepsilon,z)) + (1 - \gamma) \mathbb{E}_{\tilde{p}_2}(\lambda(\varepsilon,z)).
		\end{equation}
	By considering individually the behaviour near $\varepsilon = 0$ and $\varepsilon = 1$, we can see that
		\begin{equation*}
		\begin{aligned}
			\lambda(\varepsilon,z) &\sim \lambda(0,z) + \lambda'(0,z) \varepsilon && %
			\Rightarrow &\mathbb{E}_{\tilde{p}_1}(\lambda(\varepsilon,z)) &\sim \lambda(0,z) + \lambda'(0,z) E_1, \\
			\lambda(\varepsilon,z) &\sim \lambda(1,z) - \lambda'(1,z)(1 - \varepsilon)&& %
			\Rightarrow &\mathbb{E}_{\tilde{p}_2}(\lambda(\varepsilon,z)) &\sim \lambda(1,z) - \lambda'(1,z) E_2.
		\end{aligned}
		\end{equation*}
	Finally, we can determine $\gamma$ in terms of $E_1$ and $E_2$ by considering that the \textit{net transfer of phenotypes} from near $\varepsilon = 0$ (and similarly in the reverse direction) is still given by $\omega(1-q)$, while the order of integration and differentiation can be exchanged to see that the net growth of phenotypes near $\varepsilon = 0$ is simply given by $\mathbb{E}_{\tilde{p}_1}(\lambda(\varepsilon))$. Thus, we can use the result from \Cref{qssbinary} to see that the QSS proportion of cells near the dormant state is given by
		\begin{equation}\label{gamma}
			\gamma = \dfrac{-a + b + \Delta\lambda + \sqrt{4b\lambda + (-a + b + \Delta\lambda)^2}}{2\Delta\lambda},
		\end{equation}
	for $a = \omega (1 - q)$, $b = \omega q$ and where $\Delta\lambda = \mathbb{E}_{\tilde{p}_2}(\lambda(\varepsilon)) - \mathbb{E}_{\tilde{p}_1}(\lambda(\varepsilon))$ (such that we consider that, equivalently, the net growth rate of the dormant population to be zero; results relating to the population distribution are insensitive to such growth rate translation, or equivalently, scaling at the population level).
	
	 To determine asymptotic expressions for $E_1$ and $E_2$, we first substitute \cref{ptildemix} into \cref{qss1}, to obtain
		\begin{equation}\label{ptilde1}
			0 = (\gamma \tilde{p}_1 + (1 - \gamma)\tilde{p}_2)\Big[\lambda - \gamma \mathbb{E}_1(\lambda) - (1 - \gamma)\mathbb{E}_2(\lambda)  - \omega\Big] + \omega (q p_1 + (1 - q)p_2),
		\end{equation}
	Recalling that \cref{ptilde1} must apply for all $\varepsilon$, we next assume that the density of $\tilde{p}_1$ and $p_1$ is negligible near $\varepsilon = 1$, and similar for $\tilde{p}_2$ and $p_2$ near $\varepsilon = 0$. Following several substitutions and by considering behaviour up to $\mathcal{O}(\eta_i)$, we obtain
		\begin{subequations}
		\begin{align*}
			0 &= \sqrt{\dfrac{2}{\pi}} q\omega \eta_1 - E_1 \gamma\big[\omega + (1 - \gamma)\big(\lambda(1,z) - \lambda(0,z)\big)\Big],\\
			0 &= \sqrt{\dfrac{2}{\pi}} (1 - q) \omega \eta_2 - E_2(1 - \gamma)\Big[\omega - \gamma\big(\lambda(1,z) - \lambda(0,z)\big)\Big].
		\end{align*}
		\end{subequations}
	It remains to substitute the expression for $\gamma$ (\cref{gamma}), and consider matched order terms to obtain asymptotic expressions for $E_1$ and $E_2$. Substituting the result back into \cref{Eplambda}, we obtain
		\begin{equation}
			\mathbb{E}_{\tilde{p}}(\lambda(\varepsilon,z)) = c_0(z) + c_1(z) \eta_1 + c_2(z) \eta_2,
		\end{equation}
	where
		\begin{subequations}
		\begin{align*}
			c_0(z) &= \dfrac{1}{2}\Big(\Delta(z) - \omega + \sqrt{\Omega}\Big),\\
			c_1(z) &= \dfrac{\lambda'(0,z)\Big(\big[(2q-1) \omega - \Delta(z)\big] + 1\Big)}{\sqrt{2\pi\Omega(z)}},\\
			c_2(z) &= \dfrac{\lambda'(1,z)\Big(\big[(2q-1) \omega - \Delta(z)\big] - 1\Big)}{\sqrt{2\pi\Omega(z)}},
		\end{align*}
		\end{subequations}
	for
		\begin{align*}
			\Delta(z) &:= \lambda(1,z) - \lambda(0,z),\\
			\Omega(z) &:= (\omega - \Delta(z))^2 + 4(1-q) \omega \Delta(z),
		\end{align*}
	and where $\lambda'(\varepsilon,z)$ indicates a derivative with respect to $\varepsilon$. 
	
	An asyptotic expression for the fitness is, therefore, given by
		\begin{equation*}
			\varphi \sim C_0 + C_1 \eta_1 + C_2 \eta_2,
		\end{equation*}
	where
		\begin{equation*}
			C_i = \int_{-\infty}^\infty c_i(z) \Phi'(z)\,\dd z.	
		\end{equation*}
	%

	\subsection{Fluctuations in the fittest phenotype}
	
	We consider that the growth rate is given by (for now, general) $\lambda(\varepsilon,w)$, where the environment, $w$, fluctuates according to an Ornstein-Uhlenbeck process with equilibrium distribution $w \sim \mathcal{N}(0,\sigma)$. For simplicity, we consider that $\lambda(\varepsilon,w)$ is symmetric about $\varepsilon$, such that the optimal phenotype in a homogeneous strategy is given by $\varepsilon^* = 0$ (in the main document, we translate the phenotype index such that $\varepsilon^* = \tfrac{1}{2}$ so that the phenotype index remains bounded $\varepsilon \in [0,1]$). 
	
	The target distribution is given by $\hat{p}(\varepsilon) = \mathcal{N}(0,\eta)$, for $\eta \ll 1$. For a fixed value of the environment (i.e., in the QSS regime), in the vicinity of $\varepsilon^* = 0$ we approximate
		\begin{equation}\label{fit_lambda}
			\lambda(\varepsilon,w) \sim \lambda(0,w) + \lambda'(0,w)\varepsilon + \dfrac{\lambda''(0,w)}{2}\varepsilon^2,
		\end{equation}
	where $\lambda'(\varepsilon,w)$ denotes a derivative with respect to $\varepsilon$. 
	
	Next, we substitute \cref{fit_lambda} into the governing equation for $\tilde{p}$ (\cref{qss1}), multiply by $\varepsilon^n$, and integrate to yield
		\begin{equation}
			0 = \lambda'(0,w) E_{n+1} + \tfrac{1}{2}\lambda''(0,w) E_{n+2} - \lambda'(0,w) E_1E_n - \tfrac{1}{2}\lambda''(0,w) E_2E_n - \omega E_n + \omega M_n,
		\end{equation}
	where we have defined
		\begin{equation}
			E_n = \mathbb{E}_{\tilde{p}}(\varepsilon^n) \quad\text{and}\quad M_n = \mathbb{E}_{\hat{p}}(\varepsilon^n).
		\end{equation}

	If we proceed as before and assume that $E_1 \sim \mathcal{O}(\eta)$, we find that the equation can only be satisfied at $\mathcal{O}(\eta)$ through the trivial solution $E_1 = 0$ that later leads to a contradiction. Thus, we progress to the next order and assume that potentially both $E_1,E_2 \sim \mathcal{O}(\eta^2)$, while $E_n \sim \mathcal{O}(\eta^3)$ for $n \ge 3$. For $n = 1$ and $n = 2$, we obtain up to order $\mathcal{O}(\eta^2)$:
		\begin{subequations}
		\begin{align*}
			0 &= \lambda'(0,w) E_2 - \omega E_1,\\
			0 &= -\omega E_2 + \omega \eta^2,
		\end{align*}
		\end{subequations}
	such that $E_1 = \lambda'(0)\omega^{-1}/\eta^2$ and $E_2 = \eta^2$. Substituting back into \cref{fit_lambda}, we find that
		\begin{equation}\label{fit_lambda}
			\mathbb{E}_{\tilde{p}}(\lambda(\varepsilon,w)) \sim c_0(w) + c_2(w) \eta^2,
		\end{equation}
	where the first order correction vanishes, and
		\begin{subequations}
		\begin{align*}
			c_0(w) &= \lambda(0,w),\\
			c_2(w) &= \dfrac{\lambda'(0,w)^2}{\omega} + \dfrac{\lambda''(0,w)}{2}.
		\end{align*}
		\end{subequations}

	As in the previous section, the fitness is thus given by
		\begin{equation*}
			\varphi \sim C_0 + C_2 \eta_2,
		\end{equation*}
	where
		\begin{equation}\label{fit_Ci}
			C_i = \int_{-\infty}^\infty c_i(z) \phi_w(w)\,\dd w,
		\end{equation}
	where $\phi_w(w)$ is the density function for the random variable $w$. 
	
	For the specific formulation examined in the main document
		\begin{equation*}
			\lambda(\varepsilon,w) = k \exp\left(\dfrac{-(\varepsilon - w)^2}{\xi^2}\right),
		\end{equation*}
	we can compute the integral in \cref{fit_Ci} analytically to produce the expression given in the main document.
